\documentclass[12pt]{article}
\usepackage{amsmath,amsfonts,amssymb}
\usepackage[english]{babel}
\usepackage{graphicx}
\newcommand{\llra}{\leftrightarrow}
\newcommand{\ra}{\rightarrow}

\newcommand{\lra}{\longrightarrow}
\newcommand{\Lra}{\Longrightarrow}
\newcommand{\ua}{\uparrow}
\newcommand{\da}{\downarrow}
\newcommand{\half}{{\textstyle{\frac{1}{2}}}}
\newcommand{\third}{{\textstyle{\frac{3}{2}}}}

\newcommand{\be}{\begin{equation}}
\newcommand{\ee}{\end{equation}}
\newcommand{\bea}{\begin{eqnarray}}
\newcommand{\eea}{\end{eqnarray}}

\begin{document}

\pagestyle{empty}
\bigskip

\bigskip

\begin{center}

\vfill

{\Large \textbf{\textsf{Symmetry and Minimum Principle at the      
                    Basis \\ \vspace{2mm} of the Genetic Code}}}\footnote{Based on talks given at:  BelBI2016 International Symposium, University of Belgrade, Serbia \\  and   BIOMAT 2016 International Symposium, Nankai University, Tianjin,China}

\vspace{8mm}

{\large   A. Sciarrino}

\vspace{4mm}

 \emph{I.N.F.N., Sezione di Napoli  \\ Complesso Universitario di Monte S. Angelo \\ Via Cinthia, I-80126 Napoli, Italy \\ nino.sciarrino@gmail.com}
 
 \vspace{8mm}

{\large   P.Sorba}

\vspace{4mm}

 \emph{LAPTH,Laboratoire d'Annecy-le-Vieux de Physique Th\' eorique CNRS \\ Universit\' e de Savoie \\
 Chemin de Bellevue, BP 110,\\
 F-74941 Annecy-le-Vieux, France \\
E-mail: paul.sorba@lapth.cnrs.fr}

\vspace{10mm}

{\bf To appear in BIOMAT 2016, 326 - 362, 2017} 

\vspace{2mm}

\end{center}

\vspace{4mm}

\begin{abstract}
The importance of the notion of symmetry in physics is well established: could it also be the case for the genetic code? In this spirit, a model for the Genetic Code based on continuous symmetries and entitled the  ``Crystal Basis Model" has been proposed a few years ago. The present paper is a review of the model, of some of its first applications as well as of its recent developments. Indeed, after a  motivated presentation of our mathematical model, we  illustrate its pertinence by applying it for the elaboration and verification of sum rules for codon usage probabilities, as well as for establishing relations and some predictions between physical-chemical properties of amino-acids.  Then, defining in this context a ``bio-spin" structure for the nucleotides and codons, the interaction between a couple of codon-anticodon can simply be represented by a (bio) spin-spin potential. This approach will constitute the second part of the paper where, imposing the minimum energy principle, an analysis of the evolution of the genetic code can be performed with good agreement with the generally accepted scheme. A more precise study of this interaction model provides informations on codon bias, consistent with data.
\end{abstract}

\vspace{2mm}

\newpage

\pagestyle{plain}
\setcounter{page}{1}

\section{Introduction}

The sciences of life offer an important domain of investigations for the physicist. Already about seventy years ago, Erwin Schr\"odinger provided in his book  ``{\it What is life ?}''  \cite{Schr} some ideas about the possible role of a  ``new physics''  in this domain, imagining for example mutations to be directly linked to quantum leads. As can be read there: 

 {\em`` living matter, while not eluding the  ``laws of physics''   as established up to date, is likely to involve  ``other laws of physics"  hitherto unknown, which however, once they have been revealed, will form just as integral a part of science as the former''}.

Among the mathematical tools which played in the second part of the twentieth century and are still playing an essential role in theoretical physics, and in particular in particle physics, is the one of Group theory, this concept being usually called in physics {\em Symmetry}, or {\em Invariance}. It is this notion which is at the basis of our model for describing the genetic code and developing a theoretical approach of its biological properties. 

The idea of symmetry, or invariance, can be used in different ways, but to illustrate the one we need today, let us take an example. Consider an electron $e^{-}$. An important physical quantity attached to it is its spin. And actually, as you know, there are two states for the spin of the electron, called up and down, or + and - (or +1/2 and  -1/2 following the notation you choose, and we use to say that the spin of $e^{-}$ is 1/2). Mathematically, these two states can be seen as orthogonal vectors of the 2-dim complex Euclidean space\footnote{Such a space is denoted as a 2-dim Hilbert space.}, on which acts the group of 2 by 2 unitary matrices, called $\mathcal {SU}(2)$, transforming one state into another one. It is a Lie group and considering its Lie algebra, there exists an element on it, a $2  \times 2$ matrix with eigenvalues +1/2 and -1/2 associated to the eigenvectors which are the states up and down. If you consider a vector boson, (e.g. the $W$ boson which mediates the weak interaction) there are three states of spin, denoted +1, 0, -1 and we can represent the elements of the group $\mathcal {SU}(2)$ by $3 \times 3$ matrices acting on a 3-dim  Hilbert space. It is this notion that we will use to construct our model describing the genetic code. 

At this point, let us mention two essential aspects of genetics on which we propose to use our model: the DNA structure on one hand and the mechanism of polypeptide fixation from codons on the other hand. But, it might be good to start by reminding some essential features on the genetic code.
First, as well known, the DNA macromolecule is constituted by two chains of nucleotides  wrapped in a double helix shape. There are four different nucleotides,
characterized by their bases: adenine (A) and guanine (G) deriving from purine, and cytosine (C) and thymine (T) coming from pyrimidine. Note also the A (reps. T) base in one strand
is connected with two hydrogen bonds to a T (resp. A) base in the other strand, while a C (resp. G) base is related to a G (reps. C) base with three hydrogen bonds. The genetic information is transmitted to the cytoplasm via the messenger ribonucleic acid (mRNA). During this operation, called transcription, the A, G, C, T bases in the DNA are associated respectively  to the U, C, G, A bases, U denoting the uracile base. Then it will be through a ribosome that a triplet of nucleotides or codon will be related to an amino acid (a.a.).  More precisely, a codon is defined as an ordered sequence of three nucleotides, e.g. AAG, ACG, etc., and one enumerates in this way $ 4 \times 4 \times 4 = 64$
different codons. In the universal eukariotic code (see Table \ref{table:gc}), 61 of such triplets can be connected in an unambiguous way to the amino-acids, except the three following triplets UAA, UAG and UGA, which are called non-sense or stop  codons, the role of which is to stop the biosynthesis. Indeed the genetic code is the association between codons and amino-acids. But since one distinguishes only 20 amino-acids \footnote{Alanine (Ala), Arginine (Arg), Asparagine (Asn), 
Aspartic acid (Asp), Cysteine (Cys), Glutamine (Gln), Glutamic acid (Glu), 
Glycine (Gly), Histidine (His), Isoleucine (Ile), Leucine (Leu), Lysine 
(Lys), Methionine (Met), Phenylalanine (Phe), Proline (Pro), Serine (Ser), 
Threonine (Thr), Tryptophane (Trp), Tyrosine (Tyr), Valine (Val).} related to the 61 codons, it follows that the genetic code is degenerate. Still considering the standard eukariotic code, one observes sextets, quadruplets, triplet, doublets and singlet of codons, each multiplet corresponding to a specific amino-acid (a.a.).

In the mathematical framework we have proposed \cite{FSS98}, the codons appear as composite states of nucleotides. More precisely, the codons are obtained as tensor products of nucleotides, the four nucleotides being assigned to the fundamental representation of the quantum group  ${\mathcal U_{q}}(sl(2) \oplus sl(2)) $  in the limit of the deformation parameter  $q \to 0$. The use of a quantum group  in the limit $q \to 0$ is essential to take into account the nucleotide ordering (see Table 1).  Of course, the reader who is not interested in the mathematical aspects can jump over them and focus his attention on the biophysical results which are presented hereafter. However, for the reader who wishes to better understand our approach, we have devoted a rather developed    ``tutorial''  on group theory at the end of this review. The first part of this appendix deals with general notions and properties of Lie groups while the second part shows explicitly how are constructed the codons as representation states of the quantum group above mentioned. 

We have distinguished two parts in this review. 

The first one starts with a rapid recalling of the main aspects of our model that  we called  ``Crystal Basis Model''. It is followed by two examples of applications. The first one concerns the setting of sum rules for codon usage probabilities \cite{FSS03}: it is deduced that the sum of usage probabilities of codons with C and A in the third position for the quartets and/or sextets is independent of the biological species for vertebrates. The second application deals with the physical-chemical properties of amino-acids for which a set of relations have been derived and compared with the experimental data \cite{FSS02}. A prediction for the not yet measured thermo-dynamical parameters of three amino-acids is also proposed.

Another important notion in physics is the principle of minimal action, or use of minimum of energy. This is the second main idea that we will keep with us in the second part of this review in which a codon-anticodon interaction potential is proposed \cite{SS12},  still in the framework of the  ``Crystal Basis Model''.  Such a study will first allow to determine the structure of the minimum set of 22 anticodons allowing the translational-transcription for animal mitochondrial code. The results are in very good agreement with the observed anticodons. Then, the evolution of the genetic code is considered, with 20 amino-acids encoded from the beginning, from the viewpoint of codon-anticodon interaction. Following the same spirit as above, a determination of the structure of the anticodons in the Ancient, Archetypal and Early Genetic codes is obtained \cite{SS13}. Most of our results agree with the generally accepted scheme. Finally, keeping still at hand the minimization of our codon-anticodon interaction potential, codon bias are discussed, providing  inequalities between codon usage probabilities for quartets of codons \cite{SS16}.  Performing this study separately for the Early and for the Eukariotic genetic code, we observe a consistency with the obtained results as well as good agreement with the available data. Last but not least, an analysis of the coherent change of sign, in the evolution from the Early to the Eukaryotic code, of the two parameters regulating our interaction potential is performed. 

Some general remarks are gathered in the conclusion, while, as already mentioned, a large appendix is devoted to the mathematical aspect of symmetry.

As this paper is essentially a review of the Crystal Basis Model, we have limited the references and only provided those directly connectedÊ to our approach. The interested reader can find in each quoted paper the relative biography.

\section{PART 1:  Crystal basis model and application}

\subsection{A group theoretical model of the genetic code}
\label{sect:model}

We consider the four nucleotides as basic states of the $(\half,\half)$ 
representation of the \, ${\mathcal U}_{q}(sl(2) \oplus sl(2))$ quantum enveloping 
algebra in the limit $q \to 0$ \cite{FSS98}. A triplet of nucleotides will then be 
obtained by constructing the tensor product of three such four-dimensional 
representations. Actually, this approach mimicks the group theoretical 
classification of baryons made out from three quarks in elementary 
particles physics, the building blocks being here the A, C, G, T/U 
nucleotides. The main and essential difference stands in the property of a 
codon to be an \emph{ordered} set of three nucleotides, which is not the 
case for a baryon.

Constructing such pure states is made possible in the framework of any 
algebra ${\mathcal U}_{q \to 0}({\mathcal G})$ with ${\mathcal G}$ being any 
(semi)-simple classical Lie algebra owing to the existence of a special 
basis, called crystal basis, in any (finite dimensional) representation of 
${\mathcal G}$. The algebra ${\mathcal G} = sl(2) \oplus sl(2)$ appears the most 
natural for our purpose. The complementary rule in the DNA--mRNA 
transcription may suggest to assign a \emph{quantum number} with opposite 
values to the couples (A,T/U) and (C,G). The distinction between the purine 
bases (A,G) and the pyrimidine ones (C,T/U) can be algebraically 
represented in an analogous way. Thus considering the fundamental 
representation $(\half,\half)$ of $sl(2) \oplus sl(2)$ and denoting $\pm$ 
the basis vector corresponding to the eigenvalues $\pm\half$ of the $J_{3}$ 
generator in any of the two $sl(2)$ corresponding algebras, we will assume 
the following ``biological'' spin structure:

\begin{eqnarray}
\label{eq:gc1}
&sl(2)_{H}& \nonumber \\
C \equiv (+,+) &\qquad\longleftrightarrow\qquad& U \equiv (-,+) 
\nonumber \\
\Bigg. sl(2)_{V} \updownarrow && \updownarrow sl(2)_{V} \\
\Bigg. G \equiv (+,-) &\qquad\longleftrightarrow\qquad& A \equiv (-,-) 
\nonumber \\
&sl(2)_{H}& \nonumber
\end{eqnarray}
the subscripts $H$ (:= horizontal) and $V$ (:= vertical) being just 
added to specify the algebra.

Now, we consider the representations of ${\mathcal U}_{q}(sl(2))$ and more 
specifically the crystal bases obtained when $q \to 0$. Introducing in 
${\mathcal U}_{q \to 0}(sl(2))$ the operators $J_{+}$ and $J_{-}$ after 
modification of the corresponding simple root vectors of ${\mathcal 
U}_{q}(sl(2))$, a particular kind of basis in a ${\mathcal 
U}_{q}(sl(2))$-module can be defined. Such a basis is called a crystal 
basis and carries the property to undergo in a specially simple way the 
action of the $J_{+}$ and $J_{-}$ operators: as an example, for any couple 
of vectors $u,v$ in the crystal basis ${\bf B}$, one gets $u = J_{+} v$ if 
and only if $v = J_{-} u$. More interesting for our purpose is the crystal 
basis in the tensorial product of two representations. Then the following 
theorem holds \cite{K} (written here in the case of $sl(2)$):

\noindent {\bf Theorem 2.1}:
Let ${\bf B}_{1}$ and ${\bf B}_{2}$ be the crystal bases of the 
$M_{1}$ and $M_{2}$ ${\mathcal U}_{q \to 0} (sl(2))$-modules 
respectively.Then for $u \in {\bf B}_{1}$ and $v \in {\bf B}_{2}$, 
we have:
\begin{eqnarray}
&& J_{-}(u \otimes v) = \left\{
\begin{array}{ll}
J_{-} u \otimes v & \exists \, n \ge 1 \mbox{ such that } 
J_{-}^n u \ne 0 \mbox{ and } J_{+}^n v = 0 \\
u \otimes J_{-} v & \mbox{otherwise} \\
\end{array} \right.\\
&& J_{+} (u \otimes v) = \left\{
\begin{array}{ll}
u \otimes J_{+} v & \exists \, n \ge 1 \mbox{ such that } 
J_{+}^n v \ne 0 \mbox{ and } J_{-}^n u = 0 \\
J_{+} u \otimes v & \mbox{otherwise} \\
\end{array} \right.
\label{th:1}
\end{eqnarray}

Note that the tensor product of two representations in the crystal basis is 
not commutative. In the case of our model, we only need to 
construct the $n$-fold tensor product of the fundamental representation 
$(\half,\half)$ of ${\mathcal U}_{q \to 0}(sl(2) \oplus sl(2))$ by itself.

In Table \ref{table:rep} we report the assignments of the codons of the eukariotic code (the upper label denotes different irreducible 
representations) and, respectively the amino-acid content of the $\otimes^3 (\half,\half)$ representations.The codon content in each of the obtained irreducible representations is also expressed at the end of this subsection.

Let us insist on the choice of the crystal basis, which exists only in the 
limit $q \to 0$. In a codon the order of the nucleotides is of fundamental 
importance (e.g. CCU $\to$ Pro, CUC $\to$ Leu, UCC $\to$ Ser). If we want 
to consider the codons as composite states of the (elementary) nucleotides, 
this surely cannot be done in the framework of Lie (super)algebras. Indeed 
in the Lie theory, the composite states are obtained by performing tensor 
products of the fundamental irreducible representations. They appear as 
linear combinations of the elementary states, with symmetry properties 
determined from the tensor product (i.e. for $sl(n)$, by the structure of 
the corresponding Young tableaux).

On the contrary the crystal basis provides us with the mathematical 
structure to build composite states as \emph{pure} states, characterised by 
the order of the constituents. In order to dispose of such a basis, we need 
to consider the limit $q \to 0$. Note that in this limit we do not deal 
anymore either with a Lie algebra or with an universal deformed enveloping 
algebra.

To represent a codon, we have to perform the tensor product of three 
$(\half,\half)$ representations of $\mathcal {U}_{q \to 0}(sl(2) \oplus 
sl(2))$. However, it is well-known (see Table \ref{table:gc}) that in a 
multiplet of codons relative to a specific amino-acid, the two first bases 
constituent of a codon are ``relatively stable'', the degeneracy being 
mainly generated by the third nucleotide. We consider first the tensor 
product:
\be
(\half,\half) \, \otimes \, (\half,\half) = (1,1) \, \oplus \, (1,0) \, 
\oplus \, (0,1) \, \oplus \, (0,0)
\label{eq:gc4}
\ee
where inside the parenthesis, $j=0,\half,1$ is put in place of the 
$2j+1=1,2,3$ respectively dimensional $sl(2)$ representation. We get, using 
Theorem 2.1, the following tableau:
\be
\begin{array}{lcccc}
\to \,\, su(2)_H \qquad \qquad &(0,0) & (\mbox{CA}) &\qquad 
\qquad (1,0) & (
\begin{array}{ccc} 
\mbox{CG} & \mbox{UG} & \mbox{UA} \\
\end{array}) \\
\downarrow \\
su(2)_V &(0,1) & \left(
\begin{array}{c}
\mbox{CU} \\
\mbox{GU} \\
\mbox{GA} \\
\end{array}
\right) & \qquad \qquad (1,1) & \left(
\begin{array}{ccc}
\mbox{CC} & \mbox{UC} & \mbox{UU} \\
\mbox{GC} & \mbox{AC} & \mbox{AU} \\
\mbox{GG} & \mbox{AG} & \mbox{AA} \\
\end{array}
\right)
\end{array}
\label{eq:dinucl}
\ee
From Table \ref{table:rep}, the dinucleotide states formed by the first 
two nucleotides in a codon can be put in correspondence with quadruplets, 
doublets or singlets of codons relative to an amino-acid. Note that the 
sextets (resp. triplets) are viewed as the sum of a quadruplet and a 
doublet (resp. a doublet and a singlet). Let us define the ``charge'' $Q$ 
of a dinucleotide state by
\be
Q = J_{H,3} + \frac{1}{4} \, C_{V} (J_{V,3}+1) - \frac{1}{4}
\label{eq:Q}
\ee
$J_{\alpha,3}$ ($\alpha = H,V$) stands for the diagonalised 
$sl(2)_{\alpha}$ generator. The operator $C_{\alpha}$ is a Casimir operator 
of ${\mathcal U}_{q \to 0}(sl(2)_{\alpha})$ in the crystal basis. It commutes 
with $J_{\alpha\pm}$ and $J_{\alpha,3}$ and its eigenvalues on any vector 
basis of an irreducible representation of highest weight $j$ is $j(j+1)$, 
that is the same as the undeformed standard second degree Casimir operator 
of $sl(2)$. Its explicit expression is
\begin{equation}
C_{\alpha} = (J_{\alpha,3})^{2} + \half \sum_{n \in \mathbf{Z_+}} \sum_{k=0}^n 
(J_{\alpha-})^{n-k} (J_{\alpha+})^n (J_{\alpha-})^k 
\label{eq:Cas}
\end{equation}

Note that for $sl(2)_{q \to 0}$ the Casimir operator is an infinite series 
of powers of $J_{\alpha\pm}$. However in any finite irreducible 
representation only a finite number of terms gives a non-vanishing 
contribution. \\
The dinucleotide states are then split into two octets with respect to the 
charge $Q$: the eight \emph{strong} dinucleotides associated to the 
quadruplets (as well as those included in the sextets) of codons satisfy $Q 
> 0$, while the eight \emph{weak} dinucleotides associated to the doublets 
(as well as those included in the triplets) and eventually to the singlets 
of codons satisfy $Q < 0$. Let us remark that by the change $ C \llra A $ 
and $ U \llra G $, which is equivalent to the change of the sign of 
$J_{\alpha,3}$ or to reflexion with respect to the diagonals of the 
eq.(\ref{eq:gc1}), the 8 strong dinucleotides are transformed into weak 
ones and vice-versa.

If we consider the three-fold tensor product, the content into irreducible 
representations of ${\mathcal U}_{q \to 0}(sl(2) \oplus sl(2))$ is given by:
\begin{equation}
\label{eq:gc5}
(\half,\half) \otimes (\half,\half) \otimes (\half,\half) = 
(\third,\third) \oplus 2 \, (\third,\half) \oplus 2 \, (\half,\third) 
\oplus 4 \, (\half,\half)
\end{equation}
The structure of the irreducible representations of the r.h.s. of eq. 
(\ref{eq:gc5}) is (the upper labels denote different irreducible 
representations):
\begin{eqnarray*}
&(\third,\third) \equiv \left( 
\begin{array}{cccc} 
\mbox{CCC} & \mbox{UCC} & \mbox{UUC} & \mbox{UUU} \\
\mbox{GCC} & \mbox{ACC} & \mbox{AUC} & \mbox{AUU} \\
\mbox{GGC} & \mbox{AGC} & \mbox{AAC} & \mbox{AAU} \\
\mbox{GGG} & \mbox{AGG} & \mbox{AAG} & \mbox{AAA} \\
\end{array}
\right)& \\
&& \\
&(\third,\half)^{1} \equiv \left( 
\begin{array}{cccc} 
\mbox{CCG} & \mbox{UCG} & \mbox{UUG} & \mbox{UUA} \\
\mbox{GCG} & \mbox{ACG} & \mbox{AUG} & \mbox{AUA} \\
\end{array}
\right)& \\
&& \\
&(\third,\half)^{2} \equiv \left( 
\begin{array}{cccc} 
\mbox{CGC} & \mbox{UGC} & \mbox{UAC} & \mbox{UAU} \\
\mbox{CGG} & \mbox{UGG} & \mbox{UAG} & \mbox{UAA} \\
\end{array}
\right)& \\
&& \\
&(\half,\third)^{1} \equiv \left( 
\begin{array}{cc}
\mbox{CCU} &\mbox{UCU} \\
\mbox{GCU} & \mbox{ACU} \\
\mbox{GGU} & \mbox{AGU} \\
\mbox{GGA} & \mbox{AGA} \\
\end{array} 
\right)
\qquad \qquad
(\half,\third)^{2} \equiv \left( 
\begin{array}{cc}
\mbox{CUC} & \mbox{CUU} \\
\mbox{GUC} & \mbox{GUU} \\
\mbox{GAC} & \mbox{GAU} \\ 
\mbox{GAG} & \mbox{GAA} \\
\end{array} 
\right)& \\
&& \\
&(\half,\half)^{1} \equiv \left( 
\begin{array}{cc}
\mbox{CCA} & \mbox{UCA} \\ 
\mbox{GCA} & \mbox{ACA} \\ 
\end{array} 
\right)
\qquad \qquad
(\half,\half)^{2} \equiv \left( 
\begin{array}{cc}
\mbox{CGU} & \mbox{UGU} \\
\mbox{CGA} & \mbox{UGA} \\
\end{array}
\right)& \\
&& \\
&(\half,\half)^{3} \equiv \left( 
\begin{array}{cc}
\mbox{CUG} & \mbox{CUA} \\ 
\mbox{GUG} & \mbox{GUA} \\ 
\end{array} 
\right)
\qquad \qquad
(\half,\half)^{4} \equiv \left( 
\begin{array}{cc}
\mbox{CAC} & \mbox{CAU} \\ 
\mbox{CAG} & \mbox{CAA} \\ 
\end{array} 
\right)&
\end{eqnarray*}

\subsection{Applications}

\subsubsection{Sum rules of codon usage probabilities}
\label{subsect:sumrules}

Let $XZN$ be a codon in a multiplet encoding an amino acid, 
where the labels $X,Z,N$ stands for any of the four bases 
$A,C,G,U/T$. We define the relative frequency of usage of the codon $XZN$ 
as the ratio between the number of times $n_{XZN}$ the codon $XZN$ is used 
in the biosynthesis of the amino acid,  and the total number $n_{tot}$ of 
synthesised amino acid,.  Then, the frequency of usage of a codon in a 
multiplet is connected, in the limit of \emph{very large $n_{tot}$}, to its 
probability of usage $P(XZN)$:
\begin{equation}
\label{eq:proba}
P(XZN) = \lim_{n_{tot} \to \infty} \;\;\; 
\frac{n_{XZN}}{n_{tot}}
\end{equation}
with the normalization
\be
P(XZA) + P(XZC) + P(XZG) + P(XZU) = 1
\label{eq:norm}
\ee
The pattern of codon usage varies between species and even among tissues 
within a species.  Most of the analyses of the codon 
usage frequencies have adressed to analyze the relative abundance of 
specified codons in different genes of the same biological species or in 
the comparison of the relative abundance in the same gene for different 
biological species. No attention, at our knowledge, has been paid to 
analyse codon usage frequency summed over the whole available sequences to 
infer global correlations between different biological species.  

The aim of the paper \cite{FSS03} was to investigate this aspect and to 
predict a general law which should be satisfied by all the biological 
species belonging to vertebrates. 

From the definition of the usage probability for a codon $XZN$, see eq. 
(\ref{eq:proba}), it follows that our analysis and predictions hold for 
biological species with large enough statistics of codons. In the crystal 
basis model of the genetic code, each codon $XZN$ is described by a state 
belonging to an irreducible representation denoted $(J_{H},J_{V})^\xi$ ($\xi$ specifying the representation) of 
the algebra ${\mathcal U_{q}}( sl(2)_{H} \oplus sl(2)_{V})$ in the limit $q 
\to 0$. It is natural in this model to write the usage probability as a 
function of the biological species (b.s.), of the particular amino-acid and 
of the labels $J_{H}$, $J_{V}$, $J_{H,3}$, $J_{V,3}$ describing the state 
$XZN$. 

Assuming  the dependence of the amino-acid to be completely 
determined by the set of labels $Js$,  we write
\begin{equation}
\label{eq:2}
P(XZN) = P(b.s.; J_{H}, J_{V}, J_{H,3}, J_{V,3}) 
\end{equation}
Let us now make the hypothesis that we can write the r.h.s. of eq. 
(\ref{eq:2}) as the sum of two contributions: a universal function $\rho$ 
independent on the biological species at least for vertebrates and a b.s. 
depending function $f_{bs}$, i.e.
\begin{equation}
\label{eq:3}
P(XZN) = \rho^{XZ}(J_{H}, J_{V}, J_{H,3}, J_{V,3}) \; + \; 
f_{bs}^{XZ}(J_{H}, J_{V}, J_{H,3}, J_{V,3})
\end{equation}
From the analysis of the available data, we assume that
 the 
contribution of $f_{bs}$ is not negligible but could be smaller than the 
one due to $\rho$. As each state describing a codon is labelled by the 
\emph{quantum} labels of two commuting $sl(2)$, it is reasonable, at first 
approximation, to assume
\begin{equation}
\label{eq:4}
f_{bs}^{XZ}(J_{H}, J_{V}, J_{H,3}, J_{V,3}) \approx F_{bs}^{XZ}(J_{H};J_{H,3}) 
\, + \, G_{bs}^{XZ}(J_{V};J_{V,3})
\end{equation}
Now, let us analyse in the light of the above considerations the usage 
probability for the quartets Ala, Gly, Pro, Thr and Val and for the quartet 
sub-part of the sextets Arg (i.e. the codons of the form CGN), Leu (i.e. 
CUN) and Ser (i.e. UCN). \\
For Thr, Pro, Ala and Ser we can write, using Table \ref{table:rep} and 
eqs. (\ref{eq:2})-(\ref{eq:4}), with $N = A,C,G,U$,
\bea
\label{eq:5}
&& P(NCC) + P(NCA)  = \nonumber \\
&&  \rho_{C+A}^{NC} + F_{bs}^{NC}(\third;x) + 
G_{bs}^{NC}(\third;y) + F_{bs}^{NC}(\half;x') + G_{bs}^{NC}(\half;y')
\eea 
where we have denoted by $\rho_{C+A}^{NC}$ the sum of the contribution of 
the universal function (i.e. not depending on the biological species) 
$\rho$ relative to $NCC$ and $NCA$, while the labels $x,y,x',y'$ depend on 
the nature of the first two nucleotides $NC$, see Table \ref{table:rep}. 
For the same amino acid we can also write
\bea
\label{eq:6}
&&P(NCG) + P(NCU) =  \nonumber \\
&& \rho_{G+U}^{NC} + F_{bs}^{NC}(\third;x) + 
G_{bs}^{NC}(\third;y) + F_{bs}^{NC}(\half;x') + G_{bs}^{NC}(\half;y')
\eea
Using the results of Table \ref{table:rep}, we can remark that the 
difference between eq. (\ref{eq:5}) and eq. (\ref{eq:6}) is a quantity 
independent of the biological species,
\begin{equation}
\label{eq:7}
P(NCC) + P(NCA) - P(NCG) - P(NCU) \;=\; \rho_{C+A}^{NC} - \rho_{G+U}^{NC} 
\;=\; \mbox{Const.}
\end{equation}
In the same way, considering the cases of Leu, Val, Arg and Gly, we obtain 
with $W = C,G$
\begin{eqnarray}
\label{eq:8} 
P(WUC) + P(WUA) - P(WUG) - P(WUU) &=& \rho_{C+A}^{WU} - \rho_{G+U}^{WU} 
\;=\; \mbox{Const.}\nonumber \\
\label{eq:9} 
P(CGC) + P(CGA) - P(CGG) - P(CGU) &=& \rho_{C+A}^{CG} - \rho_{G+U}^{CG} 
\;=\; \mbox{Const.}  \nonumber \\
\label{eq:10}
P(GGC) + P(GGA) - P(GGG) - P(GGU) &=& \rho_{C+A}^{GG} - \rho_{G+U}^{GG} 
\;=\; \mbox{Const.} \nonumber \\
\end{eqnarray}
Since the probabilities for one quadruplet are normalised to one, from eqs. 
(\ref{eq:6})-(\ref{eq:10}) we deduce that for all the eight amino acids the 
sum of probabilities of codon usage for codons with last A and C (or U and 
G) nucleotide is independent of the biological species, i.e.
\begin{eqnarray}
\label{eq:11} 
P(XZC) + P(XZA) = \mbox{Const.} \qquad (XZ = NC, CU, GU, CG, GG) 
\end{eqnarray}
Moreover, assuming that for sextets the functions $F$ and $G$ depend really 
on the nature of the encoded amino acid rather than on the dinucleotide, we 
derive in a completely analogous way as above that for the amino acid Ser 
the sum $P'_{C+A}(S) = P(UCA) + P(AGC)$ is independent of the biological 
species. Note the that we normalize to 1 the probabilities of a quartet in a sextet.

A statistical discussion of the sum rules, in the more general context of correlations between 
the probabilities $P(XZN)$, can be found in \cite{FSS05}. 

An analysis with more recent data for more biological species can be found in \cite{CScia}.

 \subsection{Physico-chemical properties of amino-acids:relations and predictions}
 \label{subsect:predictions}

It is a known observation that a relationship exists between the 
codons and the physical-chemical properties of the coded amino acids. The 
observed pattern is read either as a relic of some kind of interaction 
between the amino acids and the nucleotides at an early stage of evolution 
or as the existence of a mechanism relating the properties of codons with 
those of amino acids. 

It is also observed that the relationship depends 
essentially on the nature of the \emph{second} nucleotide in the codons and 
it holds when the second nucleotide is A, U, C, not when it is G. 
To our 
knowledge neither the anomalous behaviour of G nor the existence of a 
closest relationship between some of the amino acids is understood. In  \cite{FSS02}
 we provided an explanation of both these facts in the framework of 
the crystal basis model of the genetic code.

\subsubsection{Relationship between the physical-chemical properties of amino 
acids}
 
We assume that some physical-chemical property of a given amino acid are 
related to the nature of the codons, in particular they depend on the 
following mathematical features, written in hierarchical order:
\begin{enumerate}
\item
the irreducible representation of the dinucleotide formed by the first two 
nucleotides;
\item
the sign of the charge $Q$  eq.(\ref{eq:Q}) on the dinucleotide state;
\item
the value of the third component of $J_{V,3}$ inside a fixed irreducible 
representations for the dinucleotides;
\item
the upper label(s) of the codon irreducible representation(s);
\end{enumerate}
Not all the physical-chemical properties are supposed to follow the scheme 
above; some of them are essentially given by the specific chemical 
structure of the amino acid itself. In the following, we analyse the 
physical-chemical properties of the amino acids in the light of the 
dinucleotide content of the irreducible representations of eq. 
(\ref{eq:dinucl}).

-- Representation (0,0): the codons of the form CAN (N = C, U, G, A) all 
belong to the irreducible representation $(\half, \half)^{4}$ and code for 
His and Gln, both being coded by doublets and differing by the value of 
$J_{V,3}$. Then we expect that the physical-chemical properties of 
His and Gln are very close.

-- Representation (1,0): we analyse the codons CG ($Q > 0$), UG, UA (both 
$Q < 0$). The codons CGS (S = C, G), resp. CGW (W = U, A), belonging to 
irreducible representation $(3/2, 1/2)^{2}$, resp. $(1/2, 1/2)^{2}$, all 
code for Arg, so we do not have any relation. The codons UGS, resp. UGW, 
belonging to irreducible representation $(3/2, 1/2)^{2}$, resp. $(1/2, 
1/2)^{2}$, code for Cys and Trp, resp. the other Cys and Ter. So we expect 
some affinity between the physical-chemical properties of Cys and Trp, not 
very strong indeed as the former is encoded by a doublet and the latter by 
a singlet. The codons UAN, belonging to the irreducible representation 
$(3/2,1/2)^{2}$, code for the Tyr and Ter. So we expect some affinity 
between the amino acids coded by UGN and UAN, in particular between Cys and 
Tyr both being coded by doublets.

-- Representation (0,1): we analyse the codons CU, GU (both $Q > 0$) and GA 
($Q < 0$). The codons CUY and GUY (Y = C, U), resp. CUR and GUR (R = G, A), 
belonging to irreducible representation $(1/2, 3/2)^{2}$, resp. $(1/2, 
1/2)^{3}$, code for Leu and Val. Therefore we do not have any relation 
between amino acids coded by the same dinucleotide, but we expect that the 
physical-chemical properties of Leu and Val are close since CU and GU both 
belong to the same irreducible representation and are both strong. The 
codons GAN belong to the irreducible representation $(1/2, 3/2)^{2}$ and 
they code Asp and Glu (both doublets). Then we expect the physical-chemical 
properties of Asp and Glu to be very close.

-- Representation (1,1): the dinucleotide irreducible representation 
($1,1$) contains five states with $Q > 0$ (CC, UC, GC, AC, GG). The codons 
CCN and UCN (resp. GCN and ACN) belong to four different irreducible 
representations and code for Pro and Ser (resp. Ala and Thr). We expect a 
strong affinity between the physical-chemical properties of Pro and Ser on 
the one hand and between the physical-chemical properties of Ala and Thr on 
the other hand. The codons GGN belong to two different irreducible 
representations and code for Gly, so we expect an affinity of 
physical-chemical properties of Gly with those of Pro, Ser, Ala, Thr. Now 
let us look at the four states with $Q < 0$ (UU, AU, AG, AA). The codons 
UUN belong to two different irreducible representations and code for Leu, 
the doublet subpart of the sextet, and for Phe (doublet). An affinity is 
expected between the physical-chemical properties of these two amino acids. 
The codons AUN belong to two different irreducible representations and code 
Ile (triplet) and Met (singlet) and, in fact, the values of 
physical-chemical properties of these two amino acids are not very 
different. The codons AGN belong to two different irreducible 
representations and code for Ser and Arg, the doublet subpart of the 
sextet, so an affinity between the physical-chemical properties of these 
codons is expected. The codons AAN belong to the same irreducible 
representation ($3/2, 3/2$) and code for Asn and Lys, so the values of the 
physical-chemical properties of these amino acids should be close.

Note that for the three sextets (Arg, Leu, Ser) the quartet (doublet) 
subpart is coded by a codon with a strong (weak) dinucleotide.

\subsubsection{Discussion}

We have compared our theoretical predictions with 10 physical-chemical 
properties:
\begin{itemize}

\item the Chou-Fasman conformational parameters 
$P_{\alpha}$, $P_{\beta}$ and $P_{\tau}$ which gives a measure of the probability of the 
amino acids to form respectively a helix, a sheet and a turn. The sum 
$P_{\alpha} + P_{\beta}$ appears more appropriate to characterise the 
generic structure forming potential and the difference $P_{\alpha} - 
P_{\beta}$ the helix forming potential, this quantity depending more on the 
particular amino acid. So we compare with $P_{\alpha}$ + 
$P_{\beta}$ and $P_{\tau}$;

\item the Grantham polarity $P_G$; 

\item the relative hydrophilicity $R_f$; 

 \item the thermodynamic activation parameters at 298 K: $\Delta H$ (enthalpy, in 
kJ/mol), $\Delta G$ (free energy, in kJ/mol) and $\Delta S$ (entropy, in 
J/mole/K); 
 
\item the negative of the logarithm of the dissociation constants at 298 K:
$pK_a$ for the $\alpha$-COOH group and $pK_b$ for the $\alpha$-NH$_3^{+}$ 
group; 
  
\item the isoelectronic point $pI$,  i.e. the $pH$ value at which no electrophoresis occurs.
 
\end{itemize}

The comparison between the theoretical relations and the experimental 
values shows:
 
($\cong$ means 
strong affinity, $\approx$ affinity, $\sim$ weak affinity):
\begin{itemize}
\item
His $\cong$ Gln -- The agreement, except for $pI$, is very good.
\item
Asp $\cong$ Glu -- The agreement, except for $P_{\tau}$, is very good.
\item
Asn $\cong$ Lys $\sim$ Arg, Ser -- The agreement, except for $pI$ and 
$P_{\tau}$ is very good. The comparison with the values of 
physical-chemical properties of Ser and Arg is satisfactory.
\item
Cys $\cong$ Tyr $\approx$ Trp -- Except for $R_f$, the agreement between 
the first two amino acids is very good, while with Trp is satisfactory.
\item
Leu $\cong$ Val -- The agreement is very good.
\item
Pro $\cong$ Ser $\approx$ Gly -- The agreement is very good, except for 
$P_{\alpha} + P_{\beta}$ and $\Delta H$, and with Gly more than 
satisfactory.
\item
Ala $\cong$ Thr $\approx$ Gly, Pro, Ser -- The agreement is very good 
between the first two amino acids except for $P_{\tau}$ and satisfactory 
with the others except for the conformational parameters.
\item
Ile $\cong$ Met $\approx$ Phe -- The agreement is very good between the 
first two amino acids and satisfactory with Phe.
\end{itemize}
So we predict that for Asp and Glu, one should find 
$\Delta H \approx 60$ kJ/mol, $-\Delta S \approx 135$ kJ/mol/K and $\Delta 
G \approx 100$ kJ/mol.

In conclusion, the values of physical-chemical properties show, with a few 
exceptions, a pattern of correlations which is expected from the 
assumptions of the crystal basis model. The remarked property that the 
amino acids coded by codons whose second nucleotide is G do not share 
similarity in the physical-chemical properties with other amino acids does 
find an explication in the model, as it is immediate to verify that there 
are no two states with G in second position which share simultaneously the 
properties of belonging to the same irreducible representation and being 
characterised by the same value of $Q$.

More details and illustrative  Tables can be found in \cite{FSS02}.


\begin{table}[hb]
\centering{\caption{The eukariotic code}}
\bigskip
{\begin{tabular}{|cc||cc||cc||cc|}
\hline
codon & a.a. & codon & a.a & codon & amino acid & codon & 
a.a. \\
\hline
CCC & Pro P & UCC & Ser S & GCC & Ala A & ACC & Thr T \\
CCU & Pro P & UCU & Ser S & GCU & Ala A & ACU & Thr T \\
CCG & Pro P & UCG & Ser S & GCG & Ala A & ACG & Thr T \\
CCA & Pro P & UCA & Ser S & GCA & Ala A & ACA & Thr T \\
\hline
CUC & Leu L & UUC & Phe F & GUC & Val V & AUC & Ile I \\
CUU & Leu L & UUU & Phe F & GUU & Val V & AUU & Ile I \\
CUG & Leu L & UUG & Leu L & GUG & Val V & AUG & Met M \\
CUA & Leu L & UUA & Leu L & GUA & Val V & AUA & Ile I \\
\hline
CGC & Arg R & UGC & Cys C & GGC & Gly G & AGC & Ser S \\
CGU & Arg R & UGU & Cys C & GGU & Gly G & AGU & Ser S \\
CGG & Arg R & UGG & Trp W & GGG & Gly G & AGG & Arg R \\
CGA & Arg R & UGA & Stop   & GGA & Gly G & AGA & Arg R \\
\hline
CAC & His H & UAC & Tyr Y & GAC & Asp D & AAC & Asn N \\
CAU & His H & UAU & Tyr Y & GAU & Asp D & AAU & Asn N \\
CAG & Gln Q & UAG & Stop  & GAG & Glu E & AAG & Lys K \\
CAA & Gln Q & UAA & Stop   & GAA & Glu E & AAA & Lys K \\
\hline
\end{tabular}}
\label{table:gc}
\end{table}
%

\newpage

\begin{table}[ht]
\centering{\caption{Assignments of the codons of the eukariotic code in the crystal basis model. The upper label denotes different irreducible 
representations.}}
{\footnotesize
\begin{tabular}{|cc|rlrr|cc|rlrr|}
\hline
codon & a.a & $J_{H}$ & $J_{V}$ & $J_{H,3}$ & $J_{V,3}$ & codon & 
a.a. & $J_{H}$ & $J_{V}$ & $J_{H,3}$ & $J_{V,3}$ \\
\hline
\bigg. CCC & Pro P & 3/2 & 3/2 & 3/2 & 3/2 & UCC & Ser S & 3/2 & 3/2 & 1/2 
& 3/2 \\
\bigg. CCU & Pro P & (1/2 & 3/2$)^1$ & 1/2 & 3/2 & UCU & Ser S & (1/2 & 
3/2$)^1$ & $-$1/2 & 3/2 \\
\bigg. CCG & Pro P & (3/2 & 1/2$)^1$ & 3/2 & 1/2 & UCG & Ser S & (3/2 & 
1/2$)^1$ & 1/2 & 1/2 \\
\bigg. CCA & Pro P & (1/2 & 1/2$)^1$ & 1/2 & 1/2 & UCA & Ser S & (1/2 & 
1/2$)^1$ & $-$1/2 & 1/2 \\
\hline
\bigg. CUC & Leu L & (1/2 & 3/2$)^2$ & 1/2 & 3/2 & UUC & Phe F & 3/2 & 3/2 
& $-$1/2 & 3/2 \\
\bigg. CUU & Leu L & (1/2 & 3/2$)^2$ & $-$1/2 & 3/2 & UUU & Phe F & 3/2 & 
3/2 & $-$3/2 & 3/2 \\
\bigg. CUG & Leu L & (1/2 & 1/2$)^3$ & 1/2 & 1/2 & UUG & Leu L & (3/2 & 
1/2$)^1$ & $-$1/2 & 1/2 \\
\bigg. CUA & Leu L & (1/2 & 1/2$)^3$ & $-$1/2 & 1/2 & UUA & Leu L & (3/2 & 
1/2$)^1$ & $-$3/2 & 1/2 \\
\hline
\bigg. CGC & Arg R & (3/2 & 1/2$)^2$ & 3/2 & 1/2 & UGC & Cys C & (3/2 & 
1/2$)^2$ & 1/2 & 1/2 \\
\bigg. CGU & Arg R & (1/2 & 1/2$)^2$ & 1/2 & 1/2 & UGU & Cys C & (1/2 & 
1/2$)^2$ & $-$1/2 & 1/2 \\
\bigg. CGG & Arg R & (3/2 & 1/2$)^2$ & 3/2 & $-$1/2 & UGG & Trp W & (3/2 & 
1/2$)^2$ & 1/2 & $-$1/2 \\
\bigg. CGA & Arg R & (1/2 & 1/2$)^2$ & 1/2 & $-$1/2 & UGA & Ter & (1/2 & 
1/2$)^2$ & $-$1/2 & $-$1/2 \\
\hline
\bigg. CAC & His H & (1/2 & 1/2$)^4$ & 1/2 & 1/2 & UAC & Tyr Y & (3/2 & 
1/2$)^2$ & $-$1/2 & 1/2 \\
\bigg. CAU & His H & (1/2 & 1/2$)^4$ & $-$1/2 & 1/2 & UAU & Tyr Y & (3/2 & 
1/2$)^2$ & $-$3/2 & 1/2 \\
\bigg. CAG & Gln Q & (1/2 & 1/2$)^4$ & 1/2 & $-$1/2 & UAG & Ter & (3/2 & 
1/2$)^2$ & $-$1/2 & $-$1/2 \\
\bigg. CAA & Gln Q & (1/2 & 1/2$)^4$ & $-$1/2 & $-$1/2 & UAA & Ter & (3/2 & 
1/2$)^2$ & $-$3/2 & $-$1/2 \\
\hline
\bigg. GCC & Ala A & 3/2 & 3/2 & 3/2 & 1/2 & ACC & Thr T & 3/2 & 3/2 & 1/2 
& 1/2 \\
\bigg. GCU & Ala A & (1/2 & 3/2$)^1$ & 1/2 & 1/2 & ACU & Thr T & (1/2 & 
3/2$)^1$ & $-$1/2 & 1/2 \\
\bigg. GCG & Ala A & (3/2 & 1/2$)^1$ & 3/2 & $-$1/2 & ACG & Thr T & (3/2 & 
1/2$)^1$ & 1/2 & $-$1/2 \\
\bigg. GCA & Ala A & (1/2 & 1/2$)^1$ & 1/2 & $-$1/2 & ACA & Thr T & (1/2 & 
1/2$)^1$ & $-$1/2 & $-$1/2 \\
\hline
\bigg. GUC & Val V & (1/2 & 3/2$)^2$ & 1/2 & 1/2 & AUC & Ile I & 3/2 & 3/2 
& $-$1/2 & 1/2 \\
\bigg. GUU & Val V & (1/2 & 3/2$)^2$ & $-$1/2 & 1/2 & AUU & Ile I & 3/2 & 
3/2 & $-$3/2 & 1/2 \\
\bigg. GUG & Val V & (1/2 & 1/2$)^3$ & 1/2 & $-$1/2 & AUG & Met M & (3/2 & 
1/2$)^1$ & $-$1/2 & $-$1/2 \\
\bigg. GUA & Val V & (1/2 & 1/2$)^3$ & $-$1/2 & $-$1/2 & AUA & Ile I & (3/2 
& 1/2$)^1$ & $-$3/2 & $-$1/2 \\
\hline
\bigg. GGC & Gly G & 3/2 & 3/2 & 3/2 & $-$1/2 & AGC & Ser S & 3/2 & 3/2 & 
1/2 & $-$1/2 \\
\bigg. GGU & Gly G & (1/2 & 3/2$)^1$ & 1/2 & $-$1/2 & AGU & Ser S & (1/2 & 
3/2$)^1$ & $-$1/2 & $-$1/2 \\
\bigg. GGG & Gly G & 3/2 & 3/2 & 3/2 & $-$3/2 & AGG & Arg R & 3/2 & 3/2 & 
1/2 & $-$3/2 \\
\bigg. GGA & Gly G & (1/2 & 3/2$)^1$ & 1/2 & $-$3/2 & AGA & Arg R & (1/2 & 
3/2$)^1$ & $-$1/2 & $-$3/2 \\
\hline
\bigg. GAC & Asp D & (1/2 & 3/2$)^2$ & 1/2 & $-$1/2 & AAC & Asn N & 3/2 & 
3/2 & $-$1/2 & $-$1/2 \\
\bigg. GAU & Asp D & (1/2 & 3/2$)^2$ & $-$1/2 & $-$1/2 & AAU & Asn N & 3/2 
& 3/2 & $-$3/2 & $-$1/2 \\
\bigg. GAG & Glu E & (1/2 & 3/2$)^2$ & 1/2 & $-$3/2 & AAG & Lys K & 3/2 & 
3/2 & $-$1/2 & $-$3/2 \\
\bigg. GAA & Glu E & (1/2 & 3/2$)^2$ & $-$1/2 & $-$3/2 & AAA & Lys K & 3/2 
& 3/2 & $-$3/2 & $-$3/2 \\
\hline
\end{tabular}
 \label{table:rep}}
\end{table}

\section{PART 2:  A  ``minimum"  principle  in the genetic code}

\subsection{ A  ``minimum"  principle in the mRNA editing}

 The  ``minimum"  principles, in their different formulations, have played and play a very relevant role in any mathematically formulated scientific theory.  The key point of a   ``minimum"  principle  is to state that an event happens along the path that  minimizes a suitable function. The mathematical formulation of a sequence in RNA or DNA in the crystal basis model allows to investigate if  some  ``minimum"  principle can be applied to the genetic code. 
 
In  \cite{FSS02b}, we have investigated the possibility to explain the position of  a nucleotide insertion in mRNA, the so called mRNA editing. The deep mechanism which causes RNA editing is still unknown.  The understanding of the event is complicated: from a thermodynamics point of view a change, i.e. C $\to$ U, takes place if it is favored in the change of entalpy or entropy, but should this be the case, the change should appear in all the organisms.  Moreover from a microscopic
(quantum mechanical) point of view, the change should occur in both directions, i.e.. C $\leftrightarrow$ U. It seems that the primary aim of
mRNA editing is the evolution and conservation of protein structures, creating a meaningful coding sequence specific for a particular amino acid sequence.

The purpose of the paper  \cite{FSS02b} was to propose an effective model to describe the RNA editing.  Our model does not explain why, where and in which organisms
editing happens, but it gives a framework to understand some specific features of the phenomenon. 
 
A consequence of the crystal basis model is that any nucleotide sequence is characterized as an element of a vector space.
Therefore, functions can be defined on this space and can be computed on the sequence of codons.  
In particular any codon is identified by a set of four
half-integer labels and functions can be defined on the codons.  We make
the assumption that the location sites for the insertion of a nucleotide
should minimize the following function for the mRNA or cDNA
\begin{equation}
\mathcal A_{0} = \exp \left[ -\sum_{k} 4 \alpha_{c} \, C_{H}^{k} \, + \, 4 \beta_{c}
\, C_{V}^{k} \, + \, 2 \gamma_{c} J_{3,H}^{k} \right]
\label{eq:A}
\end{equation}
where the sum in $k$ is over all the codons in the edited sequence,
$C_{H}^{k} $ ($C_{V}^{k} $) and $J_{3,H}^{k}$ ($J_{3,V}^{k}$), are the values of the
\emph{Casimir} operator, see eq.(\ref{eq:Cas}) and of the third component of the generator of the
$sl(2)_H$ ($sl(2)_V$), in the irreducible
representation to which the $k$-th codon belongs, see Table \ref{table:rep}. In (\ref{eq:A}) the simplified assumption that the dependence of ${\mathcal
A}_{0}$ on the irreducible representation to which the codon belongs is
given only by the values of the Casimir operators has been made.  The parameters $\alpha_{c}, \beta_{c}, \gamma_{c}$ are constants, depending on the biological species.

The minimum of $\mathcal{A}_{0}$ has to be computed in the whole set of configurations satisfying to the constraints: i) the starting point should
be the mtDNA and ii) the final peptide chain should not be modified.  It is obvious that the global minimization of expression eq.(\ref{eq:A}) is ensured
if $\mathcal A_{0}$ takes the smallest value locally, i.e. in the neighborhood of
each insertion site.  The form of the function $\mathcal {A}_{0}$ is rather
arbitrary; one of the reasons of this choice is that the chosen expression is computationally quite easily tractable.  If the parameters $\alpha_{c},
\beta_{c}, \gamma_{c}$ are strictly positive with $\gamma_{c}/6 > \beta_{c} > \alpha_{c}$, the minimization of eq.(\ref{eq:A}) explains the observed
configurations in almost  all  the considered cases,  for more details see \cite{FSS02b}.

\subsection{A  ``minimum"  principle  in the interaction codon-anticodon}
\label{subsect:min}

Given a  codon\footnote{In the paper we use the notation $N = C, A, G, U.; \; \;  R =   G, A. \;(purine)  ;\;\; Y =   C, U. \; (pyrimidine)$.} $XYZ$ ($X,Y,Z \in \{ C, A, G, U\}$) we conjecture that an anticodon    $\,X^aY^aZ^a$, where $\,Y^aZ^a=Y_cX_c$,  $N_c$ denoting  the nucleotide complementary to the nucleotide $N$ according to the Watson-Crick pairing rule\footnote{This property is observed to be verified in most, but not in all, the observed cases. To simplify we shall assume it.},
pairs to the codon $XYZ$, i.e. it is most used to ``read" the codon $XYZ$ if it minimizes the operator ${\mathcal T}$,  explicitly written in eq.(\ref{eq:T}) and computed between the 
 ``states", which can be read from Table \ref{table-mitan}, describing the codon and anticodon in the  ``crystal basis model". We write both codons (c) and anticodons (a) in $5" \to 3"$ direction. As an anticodon is antiparallel to codon, the 1st nucleotide (respectively the 3rd nucleotide) of the anticodon is paired to the 3rd (respectively the 1st) nucleotide of the codon.  
\be
{\mathcal T} = 8 c_H \,\vec{ J_H^c} \cdot  \vec{J_H^a} + 8  c_V \, \vec{J_V^c} \cdot \vec{ J_V ^a}
 \label{eq:T}
\ee
where:
\begin{itemize}
  \item $c_H.  c_V$ are constants depending on the  ``biological species" and weakly depending on the encoded a.a., as we  will later specify.
 
\item $J_H^c ,  J_V^c$  (resp. $J_H^a ,  J_V^a$) are the labels of  ${\mathcal U_{q \to 0}}(su(2)_H \oplus su(2)_V)$  specifying the state 
 
describing the codon $XYZ $ (resp. the anticodon $NY_cX_c$ pairing the codon  $XYZ$).

\item  $\vec{ J_{\alpha}^c} \cdot  \vec{J_{\alpha}^a}$ ($\alpha = H, V$) should be read as 
\be
\vec{ J_{\alpha}^c} \cdot  \vec{J_{\alpha}^a} =
\frac{1}{2}\left\{\left(\vec{J_{\alpha}^c}  \oplus \vec{J_{\alpha}}^a \right)^2 - ( \vec{J_{\alpha}^c})^2 -  ( \vec{J_{\alpha}^a})^2 \right\}
\ee
and   $\vec{J_{\alpha}^c}  \oplus  \vec{J_{\alpha}^a} \equiv \vec{J_{\alpha}^T}$  stands for the irreducible representation which the codon-anticodon state under consideration belongs to, the tensor product of   $\vec{J_{\alpha}^c}$ and $\vec{J_{\alpha}^a}$ being performed according to the rule of  \cite{K}, choosing the codon as first vector and the anticodon as second vector.  Note that 
$ \vec{J_{\alpha}}^2$ should be read as the Casimir operator whose eigenvalues are given by $
J_{\alpha}(J_{\alpha} +1 )$.
\end{itemize}

As we are interested in finding the composition of the 22  anticodons, minimun number to ensure a faihful translation,  we shall assume that the used anticodon for each quartet and each doublet is the one 
which minimizes the  averaged value of the operator  given in eq.(\ref{eq:T}), the average being performed over the 4 (2) codons  for quadruplets (doublets), see next  section.
We have found that the anticodons minimizing the conjectured operator ${\mathcal T}$ given in eq.(\ref{eq:T}), averaged  over the concerned multiplets, are in very good agreement, the results depending only on the signs of the two coupling constants,
with the observed ones, even if we have made comparison with a limited database.

The fact that the crystal basis model is able to explain, in a  relatively simple way,  the observed anticodon-codon pairing which has its roots on the stereochemical properties of nucleotides strongly suggests that our modelisation is able to incorporate some crucial features of the complex physico-chemical structure of the genetic code. Incidentally let us remark that the model  explains the symmetry codon anticodon remarked.  Let us stress that our modelisation has
a very peculiar feature which makes it very different from the standard 4-letter alphabet, used to identify the nucleotides, as well as with the usual modelisation of nucleotide chain as spin chain.  Indeed the identification of the nucleotides with the fundamental irrep. of ${\mathcal U}_{q}(su(2)_H \oplus su(2)_V)$ introduces a sort of  double ``bio-spin", which allows the description of any ordered sequence of $n$ nucleotides as as state of an irrep. and allows to describe interactions using the standard powerful mathematical language used in physical spin models.

In the  paper  \cite{SS12} we have faced the problem to find the structure of the mimimum set of anticodons and, then, we have used a very simple form for the operator ${\mathcal T}$.  
 We have not at all  discussed the possible appearance of any  other anticodon, which should require a more quantitative discussion.
For such analysis, as well as
for the eukaryotic code, the situation may be different and more than an anticodon may pair to  a quartet. 

The pattern, which in the general case may show up, is undoubtedly more complicated, depending on the biological species and on the concerned biosynthesis process, but it is natural to argue that the usage of anticodons  exhibits the general feature to assure an ``efficient"  translation process by  a number of anticodons, minimum with respect to the involved constraints.
 A more refined and quantitative analysis, which should require more data,  depends on  the value of these constants.
 
However our analysis strongly suggests that the minimum number of anticodons should be 32 (3 for the sextets, 2 for quadruplets and triplet and 1 for doublets and singlets).

In conclusion, we have found that the anticodons minimizing the conjectured operator ${\mathcal T}$ given in eq.(\ref{eq:T}), averaged  over the concerned multiplets, are in very good agreement, the results depending only on the signs of the two coupling constants,
with the observed ones, even if we have made comparison with a limited database.

The fact that the crystal basis model is able to explain, in a  relatively simple way,  the observed anticodon-codon pairing which has its roots on the stereochemical properties of nucleotides  strongly suggests that our modelisation is able to incorporate some crucial features of the complex physico-chemical structure of the genetic code. Incidentally let us remark that the model  explains the symmetry codon anticodon remarked.  Let us stress that our modelisation has
a very peculiar feature which makes it very different from the standard 4-letter alphabet, used to identify the nucleotides, as well as with the usual modelisation of nucleotide chain as spin chain.  Indeed the identification of the nucleotides with the fundamental irrep. of ${\mathcal U}_{q}(su(2)_H \oplus su(2)_V)$ introduces a sort of  double ``bio-spin", which allows the description of any ordered sequence of $n$ nucleotides as as state of an irrep. and allows to describe interactions using the standard powerful mathematical language used in physical spin models.

\begin{table}[htbp]
 \centering{ \caption{The vertebral mitochondrial code. 
 In bold (italic) the anticodons reading quadruplets (resp. doublets).}}
{\footnotesize
\begin{tabular}{|cc|c||cc|c|}
\hline
codon & a.a. & anticodon & codon & a.a. &  anticodon\\
\hline
\Big. CCC & P  & & UCC & S &  \\
\Big. CCU & P & &  UCU & S &  \\
\Big. CCG & P &{\LARGE \bf UGG} & UCG & S & {\LARGE  \bf UGA} \\
\Big. CCA & P  & &UCA & S & \\
\hline
\Big. CUC & L& & UUC & F  & \\
\Big. CUU & L & & UUU & F & {\LARGE \em GAA}\\
\Big. CUG & L & {\LARGE  \bf UAG} & UUG & L &   \\
\Big. CUA & L & &UUA & L  & {\LARGE \em UAA} \\
\hline
\Big. CGC & R &  &UGC & C & \\
\Big. CGU & R  &   & UGU & C &  {\LARGE  \em  GCA}  \\
\Big. CGG & R &  {\LARGE  \bf UCG} & UGG & W  &\\
\Big. CGA & R & &UGA & W & {\LARGE  \em UCA} \\
\hline
\Big. CAC & H &  &UAC & Y &  \\
\Big. CAU & H  & {\LARGE  \em GUG} & UAU & 
Y &  {\LARGE  \em GUA} \\
\Big. CAG & Q & &UAG & 
 {\bf Te}  & ----- \\
\Big. CAA & Q & {\LARGE  \em UUG}  &UAA &  {\bf Ter} & ----- \\
\hline
\Big. GCC & A &  & ACC & T &  \\
\Big. GCU & A & & ACU & T  & \\
\Big. GCG & A &  {\LARGE  \bf UGC} & ACG & T & {\LARGE  \bf UGU} \\
\Big. GCA & A & &ACA & T & \\
\hline
\Big. GUC & V &  & AUC & I & \\
\Big. GUU & V &  & AUU & I &    {\LARGE  \em GAU} \\
\Big. GUG & V & {\LARGE\bf UAC}& AUG & M &  \\
\Big. GUA & V& & AUA &  M &  {\LARGE  \em CAU}\\
\hline
\Big. GGC & G & & AGC & S  & \\
\Big. GGU & G & &AGU & S &  {\LARGE  \em GCU}\\
\Big. GGG & G & {\LARGE  \bf UCC} &AGG & {\bf Ter} & ---- \\
\Big. GGA & G & &AGA & {\bf Ter} & ----- \\
\hline
\Big. GAC & D & & AAC & N & \\
\Big. GAU & D &  {\LARGE  \em GUC} & AAU & N & {\LARGE  \em  GUU} \\
\Big. GAG & E &   & AAG & K  & \\
\Big. GAA & E & {\LARGE \em UUC} & AAA & K & {\LARGE  \em UUU} \\
\hline
\end{tabular}
\label{table-mitan}}
\end{table}

\subsection{The  ``minimum"  principle in the evolution of genetic code}

Using the  minimum principle stated in Subsection \ref{subsect:min} in \cite{SS13} we have analyzed
 and mathematically modellised the evolution of the genetic code in the framework on the so called  ``codon capture theoryÕÕ.

\noindent Let us  briefly summarize and   comment our results. 

We determine the structure  of the anticodons in the Ancient, Archetypal and Early  Genetic codes, that are all reconciled in a unique frame.  Most of our results  agree  with the generally accepted scheme.  Moreover the pattern of the model is surprisingly coherent.
 
The pattern of the Ancient Code  can be summarized by saying that in this primordial code the a.a.,
which would be encoded by a doublet of the type XZY are encoded by a codon ending with a C.
Similarly the a.a. which would be encoded by a doublet of the type XZR are encoded
by a codon ending with a G.

Indeed, in the Ancient Genetic Code, the sign of $c_V$ for the weak dinucleotides is undetermined,  i.e. the minimization does not depend on the sign of $c_V$. In our model, this means that there is no distinction between C (U) and G  (A). This is coherent since at this  stage there is not yet a distinction between the doublet XZR and XZY. On the contrary for strong dinucleotides  for which the role of  XZR and XZY is the same up to the  Standard Genetic Code, the sign is fixed and it does not change during the evolution.
For strong dinucleotides  and almost half of the weak ones\footnote{Let us remark that the sign of $c_H$ does not change for the weak dinucleotides which  has the value of $J_{3,H}= 0$. } there is a change in $c_H$  just when  the codon degeneracy appears, that is going from the Ancient to the Archetypal code, and the `` wobble mechanism '' is called in. For all weak  dinucleotides, the sign  of $c_V$  is  now determined and  there is a further change in the sign of $c_H$  and of $c_V$  when the  correspondence between doublets and a.a. is fixed.  

Let us remark that:
\begin{itemize}
\item for each codon there are at least two anticodons with the same value of  ${\mathcal T}$ and viceversa. This degeneracy can be removed by further terms of the interaction, not yet taken into account, but it can be also read as the ``codon disappearance''  before a new readjustment of the code.
 
\item  we remark that the presently less used (in the average)  codons,  for the a.a. encoded by doublets, are  those with last nucleotide G or U, while the most used are those with last nucleotide A or C.  So it is natural to ask the question:  why most of  the ancestral codons encoding a.a. in the Ancient Code are now  repressed ? Naively, one should expect that the ancestral codon  should be the most used one.  

\item  in our model the sign of $c_H$ for a.a. encoded by XZY in the Early Code is the same than the one in the Ancient Code, while  for a.a. encoded by XZR is the opposite.   Cys  is an exception,  but in this case the anticodon  is different in the two codes.
 This kind of argument cannot be immediately applied to a.a. encoded by quartets because in most cases the anticodon in the Archetypal,  Early or Mitochondrial Code is not the same as the one appearing in the Ancient Code and, moreover, there is an important effect due to the averaging over four codons. 
\end{itemize} 
  Analogous analysis of the codon usage frequencies for species following the Standard Code confirms generally such a pattern,  but the presence of anticodons in the Standard Code is more complicated, so we do not want to refer to these data.

Moreover, in our model naturally the anticodon with first nucleotide A does never appear, in good agreement with the observed data.

In our model we can express the evolution of the genetic code through the following pattern of the codon-anticodon interaction as 
\bea
&&  <XZN|{\mathcal T}|N^aZ^a_cX^a_c>  = \nonumber \\
&& <XZN| \left( 8 c_H \, \vec{ J_H^c} \cdot  \vec{J_H^a} + 8  c_V \, \vec{J_V^c} \cdot \vec{ J_V ^a}\right) \delta_{M^a,N^a_c}|M^aZ^a_cX^a_c> \nonumber \\
&& \Lra_{Evolution} \;\;\;  <XZN| 8 c_H \,\vec{ J_H^c} \cdot  \vec{J_H^a} + 8  c_V \, \vec{J_V^c} \cdot \vec{ J_V ^a}|M^aZ^a_cX^a_c>  \label{eq:iaa}
\eea
In the first row of eq.(\ref{eq:iaa}), the presence of the Kronecher  delta  $\delta_{M^a,N^a_c}$ enforces the Watson-Crick coupling mechanism implying $M^a = N^a_c$, while in the second row $M^a$ can be any nucleotide and the selection is implemented by the value of the operator ${\mathcal T}$, computed between the concerned states and, eventually, averaged over the multiplet taking into account the codon usage probabilities.  As example of  typical behavior of the constant $c_H$ for  weak dinucleotides, we consider the case of the AA dinucleotide:

\bea
c_H^{AAN}  > 0 \; \Lra \;  c_H^{AAN}  <  0 \;  \Lra \;  \left\{ \begin{array}{c} \;  c_H^{AAY}  > 0 \\ \\
 \; c_H^{AAR}  < 0  \end{array} \right.
\eea

The change of the sign in the coupling constants is a mathematical description to frame the modification of the  interaction codon-anticodon due to the change of the molecular structure of the nucleotides in the anticodons and of the (non local) structure of the tRNA.

Of course we have to assume that the constants $c_H$ and $c_V$ depend on the  ``time'' even if,  at this stage, only the change  of the sign in the coupling constants has been considered. 

Presumably the genetic code has not evolved along one path, It could be that  multiple branching points showed up in the course of the evolution with the advent of different genetic codes,  and then the standard genetic code would have emerged as the one exhibiting selective advantages.
For example, one can imagine that not all the  changes of the signs of 
 $c_H$ and $c_V$ would have occurred at the same time, and,  therefore, 
that several intermediate codes   would have  arisen between, say, the Ancient and the Archetypal Code. As a consequence, we believe it is more reasonable 
not to write a time-dependent evolution equation, but to verify that the existing genetic code, that is
the branching point which has survived, satisfies the required 
optimality conditions.

\subsection{The  ``minimum"  principle to explain the codon bias}

As already stated the genetic code is degenerate in the sense that a multiplet is used to encode most of the amino-acids. Some codons  in the multiplets are used much more
frequently than others to encode a particular amino-acid, i.e. there is a ``codon usage bias''. The non-uniform usage of synonymous codons is a widespread phenomenon and it is experimentally observed that the pattern of codon usage varies between species. 

The main reasons for the codon usage biases are believed to be: the genetic coding error minimization,  the CG content,  the abundance of specific anticodons in the tRNA.  
 No clear indication comes out for the existence of one or more factors  which universally  engender the codon bias, on the contrary the role of some factors is controversial. 
 
  In paper \cite{SS16}  we have analyzed  possible effects of the codon-anticodon interactions defined by the operator given in eq.(\ref{eq:T}) on the codon bias,
  according to the approach introduced in \cite{SS12},  and to propose semi-quantitative predictions of the codon bias. Moreover we briefly analyzed the codon usage bias variation along the evolution of the genetic code on the basis of the model developed in  \cite{SS13}.  In the following, we will be concerned about amino acids encoded by quartets. For the ones encoded by a sextet, that we consider as the sum of a quartet and a doublet, only the quartet will be considered. The method we developed is essentially based on the determination of the minimum values of an operator which can be seen as an interaction potential between a codon and its corresponding anticodon.
 A possible general pattern of the bias is searched   by deriving inequalities for the codon usage probabilities.  

With reference to Subsection  \ref{subsect:min} we have to minimize an expression of the type:

\bea
&& \mathcal {T}_{av}( N' YÕÕ X'' , \,XYN)= \sum _{N} \; P_N \, < N' Y'' X'' |\mathcal {T}|XYN>  \nonumber \\
&& =  P_C  \, < N' Y'' X'' | \mathcal {T}|XYC> +  \, P_U  \, < N' Y'' X'' | \mathcal {T}|XYU>   \nonumber \\
&& + P_G  \,  < N' Y'' X'' | \mathcal {T}|XYG>+ \, P_A  \,   < N' Y'' X'' | \mathcal {T}|XYA>  
\label{eq-T}
\eea
Let us recall that the expression $ < N' Y'' X'' | \mathcal {T}|XYN> $ has to be read as
 \bea
&& < N' Y'' X'' | \mathcal {T}|XYN>   \equiv  \mathcal{T} \,  \left(|XYN> \otimes \,  | N' Y'' X'' > \right) \nonumber \\
&& = \mathcal{T} \, \left( |J_H^c, J_V^c;J_{H,3}^c, J_{V,3}^c > \otimes \,  |J_H^a, J_V^a;J_{H,3}^a, J_{V,3}^a >  \right)  \nonumber \\
&& =  \lambda \,   \left( |J_H^c, J_V^c;J_{H,3}^c, J_{V,3}^c > \otimes \, |J_H^a, J_V^a;J_{H,3}^a, J_{V,3}^a >  \right) 
\eea
where we have used the correspondence
\bea
|XYN>  & \; \ra & |J_H^c, J_V^c;J_{H,3}^c, J_{V,3}^c > \nonumber \\  
  |N' Y'' X'' >  & \ra  &|J_H^a, J_V^a;J_{H,3}^a, J_{V,3}^a >
\eea
and $\lambda$ is  the eigenvalue of  $\mathcal {T}$  on the state $|J_H^c, J_V^c;J_{H,3}^c, J_{V,3}^c > \otimes \,  |J_H^a, J_V^a;J_{H,3}^a, J_{V,3}^a > $, see  \cite{SS12}  for more details.
 As the $P_{XYN}$ have to satisfy,  in addition to eq.(\ref{eq:norm}), a set of unknown constraints, we cannot impose the minimization condition in a rigorous manner, so we proceed by a heuristic  method.   Using the results of Subsection \ref{subsect:sumrules}, we are left with only two probabilities in eq.(\ref{eq-T}) and we try to argue which from the two present  $P_{XYN}$ is enhanced respect to the other one. 
For this aim we compare the two probabilities  which appear after the substitution of the other two, using eq.(\ref{eq:11}),  that have the greatest coefficient. In this way  we will get in our expression a constant terms, depending on $c_H$, $c_V$ and, generally, on $K$ ($K$ being the constant which appears in the r.h.s. of eq.(\ref{eq:11}), which has the highest  possible value, without, possibly, any specific assumption on the value of the parameters, except for the assumed sign. Then, in order to minimize the expression, it is reasonable to require that the probability with the lowest  coefficient has a higher value than the other one. Nextly, in some cases, we can derive another inequality for the complementary probabilities, according to $ K > 0,5$ or $ K < 0,5$. 

For a more detailed discussion of the difference between the minimization procedure for the Early Genetic Code  and the Eukaryotic Genetic Code,  as well as  on the assumed behavior of the coefficient $c_H$ and $c_V$  we refer to \cite{SS16}.

The  outcomes derived, which we summarize in Table \ref{th-in-q},
 are in an amazing agreement with the observed data, nevertheless the over-simplifying assumptions of our theoretical scheme and
despite that in the real world the number of operating anticodons  is greater that the minimum number  31, which implies that the matching of an a.a. encoded by a quartet is done by more than two anticodons.  Moreover  let us remark that the results found in the Early Genetic Code  survive in the Eukaryotic Genetic Code, suggesting  that we have caught some feature of a very relevant mechanism.
So we argue that codon-anticodon interaction plays a relevant role in the  codon usage bias. Moreover it seems that, in despite of its apparently fragmented behavior,  the codon bias exhibits a sort of universal feature that our approach and the Crystal Basis Model is able to take into account. 
Let us remark that, in general, for plants and bacteria and, for some extent for invertebrates, the agreement is less satisfactory. Likely this reflects the fact that the choice of the specimen for these species is too rough. The experimental data should be taken from smaller, suitably chosen, subsets of the species.

Our model seems to support the idea that the codon usage bias reflects two aspects of the tRNA population:
firstly, where there are multiple
species of tRNAs with different anticodons,  
the codons translated by the most abundant tRNA
species  are preferred; secondly,  when a tRNA can translate more than one codon, the codon best recognized by
the anticodon is preferred. In our language the most abundant tRNA and the best recognized codon are the ones which minimize the ${\mathcal T}$ operator. 
The good agreement of our results with data suggests that it may be interesting to perform a more detailed
 analysis of the two parameters controlling our codon-anticodon 
interaction potential, in a general context taking into account the complexity and the evolution  of species.
In particular, what does it mean when similar 
  parameters $c_H$ and $c_V$   conditions allow a
good prediction of codon usage for different biological sets ? For example, for Pro and Ser,
it seems that, on one hand, vertebrates and invertebrates share  similar on conditions $c_H$ and $c_V$  and, on the other hand, plants (and bacteria for Pro)
do the same.  Does the species dependent codon bias  depend on the time appearance of the amino acids ?

 Finally  there is much debate on the exact reasons for the selection of
translationally optimal codons: to increase the translational efficiency or the accuracy of the translation ? At the moment our model does not give at all  any indication for the favoured mechanism. Possibly some hints can be obtained by a study of the mutation bias along with the minimization of the codon-anticodon interaction.
 
 \begin{table}[htbp]
\centering{ \caption{Inequalities derived in the  Early  and in the Eukaryotic  Genetic Code. The value of the parameters   $c_H$ and $c_V$ is different in the two codes.} }
{\footnotesize
\begin{tabular}{|c |c | c || c | c  |}
\hline
&  \multicolumn{2}{|c||}{\large{\bf Early Code}} & \multicolumn{2}{|c|}{\large{\bf Eukaryotic Code}}  \\
 a.a. &  Parameters &   Inequalities &   Parameters &   Inequalities   \\
\hline \hline
 {\large Thr} &  ---  &  $P_C > P_G$  &$|c_H| <  3 \, |c_V|$   &  $P_C > P_U$ \\ 
 &    &    &    $|c_H| >  3 \, |c_V|$   &   $P_C > P_G$ \\ 
\hline
\hline  
{\large  Arg}&  ---  & ---  & $|c_V| < |c_H| <  2 \, |c_V|$  &   $P_C > P_G$\\  
& & &   $|c_H| <  |c_V|$ &  $P_C > P_U$ \\ 
  \hline  
 \hline
{\large  Pro} &  $|c_V| <   c_H$ & $P_A > P_U$ &  $|c_V| < 1/4 \,  |c_H|$ & $P_C > P_U$ \\ 
 &  $|c_V| >  c_H$ & $P_{A}  > P_G$   &  $|c_V| > 1/4 \,  |c_H|$&  $P_U	 > P_C$,$P_{A}  > P_G$ \\ 
 \hline  
 \hline
{\large  Leu} &  ---  & $P_G > P_C$ &  $|c_H| < 2/3 \,  |c_V|$,  & $P_U	 > P_C$  \\  
& & &  $|c_H| >  2/3 \,  |c_V|$,  & $P_C > P_U$, $P_G > P_A$    \\
  \hline\hline 
{\large  Ala} &$|c_V| <   c_H$ &    $P_U > P_A$,    $P_C > P_G$ & --- &    $P_C > P_U$ \\   
  &$|c_V| >  c_H$ &    $P_U > P_C$,    $P_A > P_G$ &   & \\   
\hline  \hline
 {\large  Gly} &  ---  &     --- & $|c_H| <   |c_V|$ &     $P_C > P_G$\\  
 & & &  $|c_H| >   |c_V|$   & $P_G  > P_C$, $P_A  > P_U$ \\
\hline    \hline
{\large  Val }  & $|c_V| < 4 c_H$   &  $P_C	 > P_G$,$P_U  > P_A$  & $|c_H| < 3 |c_V|$   &  $P_C	 > P_U$,$P_G  > P_A$ \\
& $|c_V| > 4 c_H $ & $P_C  > P_U$, $P_G  > P_A$ &  $|c_H| > 3 |c_V|$ & $P_C  > P_G$, $P_U  > P_A$ \\
\hline   \hline
{\large  Ser} &    $|c_V| <  3  c_H $ &  $P_G	 > P_C$,$P_A  > P_U$  &    $|c_V| <  \frac{5}{4} \,  |c_H|$ & $P_C > P_U$    \\
&   $|c_V| >  3  c_H $ &  $P_G	 > P_A$,$P_C  > P_U$  & $|c_V| >  \frac{5}{4} \,  |c_H|$ & $P_U  > P_C$, $P_A  > P_G$ \\
 \hline  
\end{tabular} 
 \label{th-in-q}}
 \end{table}

\section{Conclusions}

The above presented model for the genetic code can be seen as a  attempt towards a theoretical approach in the complex domain of the sciences of life. Two main notions, already contained in the title of this review, have been used:  Symmetry and Minimum Energy Principle. The first one had a particularly spectacular development during the twentieth century, as well as in mathematics under the general label of ''Group Theory'' as in several domains of fundamental physics such as Relativity, Quantum Physics and High Energy Physics. The second one, generally called the ''principle of least action'', appeared earlier with the works of Leibniz, Fermat, Euler and is generally attributed to Maupertuis, who, during the 
{\it Si\`ecle des Lumi\`eres}, felt that { ``\it Nature is thrifty in all its actions''}. 

As developed in this contribution, the Group Theory approach we proposed seems well adapted to represent the constituent actors in the genetic code and to describe some of their effects: in other words, the Crystal Basis Model allows a rather powerful ''parametrization'' of our problem. 
On the other hand, the least action principle is a more accepted and used notion in the different domains of physics, may be owing to its naturalness. In this spirit, the spin-spin like potential we have built to describe the codon-anticodon interaction can also appear  owing to its conceptual simplicity. 

For these reasons, it seems to us worthwhile to pursue investigations in the framework of our model. Such developments could be carried out as well as in the mathematical side as in the phenomenological one. Let us for example note the construction of a distance
  between two sequences of DNA or RNA \cite{Scia17} still in the framework of our model: this work deserves to be developed and applied.   In the context of evolution of the Genetic Code, it looks worthwhile to study in more details the behavior of the parameters CH and CV the role of which is determinant. More generally, a refinement of the interaction potential deserves to be considered. 

Finally, let us end this contribution by emphasizing that, among the important questions which deserve to be considered, the adaptability of the code with the increasing complexity of the organisms is a crucial one. It will be worthwhile to see to what extend our methods can be used for such a problem.  In \cite{Scia03}  a mathematical model, always in our framework, has been presented in which the main features (numbers of encoded a.a., dimensions and structure of synonymous codon multiplet) are obtained, requiring stability of the genetic code against mutations modeled by suitable operators.

Thus, our scheme appears rather well adapted  to reproduce the features of the codon capture theory as well as to provide a mathematical framework for a more quantitative and detailed description of the theory.

\section{Appendix: A short tutorial on Group Theory}

The aim of this short lesson on symmetry is to provide the necessary informations for the reader who desires to follow the mathematical aspects of group theory used in the crystal basis model developed in this review. The first part deals with general notions of a Lie group, while in the second part the notions of quantum group and so-called crystal basis are presented. 

\subsection{A I: Symmetry and Group Theory}

{\bf Definition 1}: A  group ${\mathcal G}$  is a set of elements together with a composition law - we denote it by  `` .  ''  -  such that:
\begin{enumerate}
\item $\forall \, x, y \in {\mathcal G}  \,  \quad \qquad x.y \in  {\mathcal G} \qquad  \qquad \qquad (internal \,  law) $
\item $ \forall \, x,y, z \in  {\mathcal G} \qquad (x.y).z = x.(y.z) \qquad  \;  (associativity) $
\item $ \exists \, e \in  {\mathcal G},  \; \mid  \forall \, x \in  {\mathcal G} \qquad \;\; x.e = e.x = x \qquad  (e \equiv identity) $
\item $ \forall \, x \in  {\mathcal G}, \; \mid \exists \, x^{-1} \in  {\mathcal G} \qquad x. x^{-1} =  x^{-1}.x = e   \qquad   ( x^{-1} inverse \, of \, x )$
\end{enumerate}

\noindent {\bf Examples}: $Z= \{n\}$,  $n$ integer or ${\mathbf R}= \{real  \, numbers\}$ with  `` + ''  as internal law.  $\mathbf {P_n}$: group of permutations of n objects. Set of rotations in the plane around an origin.

Actually, in Physics, groups are never considered abstractly, but as  action on some set $ \mathbf{S}$ : we will talk about group of transformations. More precisely:

\noindent {\bf Definition 2}: Let ${\mathcal G}$  be a group, $ \mathbf{S}$  a set. An action of ${\mathcal G}$   on $ \mathbf{S}$  is an application:
 $ {\mathcal G} \times  \mathbf{S} \ra  \mathbf{S}$        that is:          $(g,s) \ra g(s$)                if $g  \in {\mathcal G}$ and $ s \in \mathbf{S}$ 
such that:  $ \forall  \, g, g \prime   \in{\mathcal G}$  and $\forall \, s \in  \mathbf{S}$  :         $ g(g\prime(s))= (g . g\prime)(s)$       and  $e(s)= s$

A (transformation) group needs to be  ``represented''.

\noindent {\bf Definition 3}: A linear representation of a group ${\mathcal G}$ in a vector space $ \mathbf{V}$ (itself defined on  ${\mathbf R}$ or on  ${\mathbf C}$) is an homomorphism $D$ of  ${\mathcal G}$  on the group of linear and invertible operators on $ \mathbf{V}$, that is:
$\forall \,  g \in {\mathcal G} \ra D(g)$  such that:  $\forall g,g\prime \, \in {\mathcal G} : D(g).D(g\prime )=D(g.g\prime)$.

We note that  $D(g)$ is a $n \, \times  \, n$ matrix if $ \mathbf{V}$ is of (finite) n-dimension.

We like to say that  ``\textit{Nature is full of symmetries $\ldots$ and of symmetry breaking}'' .         For instance, let us imagine the three dimensional real space $\mathbf{R_3}$ as an homogeneous isotropic space and let us put (in an idealistic way) an electron $e^{-}$ at the point O. Then the interaction of  $e^{-}$ with a second charged particle $f$ will only depend of the distance between   $e^{-}$ and $f$:  in other words, the physics will be invariant under the three dimensional group of rotations (we denote it  $\mathcal SO(3)$, see below) around O. Now, let us introduce a magnetic field $\vec{B}$ going  in a certain direction $z$: the interaction of  $e^{-}$ with  $\vec{B}$   will no more be invariant under the whole $\mathcal {SO}(3)$ group, but only under a ``subgroup''  of it consisting of rotations in a plane perpendicular to the $z$ axis. So, we have performed a ``breaking''  of the original symmetry, and only a part of the previous set of symmetry transformations will remain as good symmetries. This phenomenon is general and it is worthwhile to know the remaining symmetries.  Whence the importance of the following - and natural - definition:

\noindent {\bf Definition 4}: A subgroup  $\mathcal{H}$ of a group $\mathcal{G}$ is a (non-empty) part $\mathcal{H} \subset \mathcal{G}$, which is a group with the composition law induced by $\mathcal{G}$.  $\mathcal{H}$ is a proper subgroup of $\mathcal{G }$ if $\mathcal{H} \neq\mathcal{H}$  and $\mathcal{H} \neq \{e\}$.

 Types of ( symmetry ) groups in physics:
\begin{itemize}
\item with a finite number of elements: case of crystallographic groups;

\item with an infinite number of elements:
\begin{itemize}
\item discrete groups (number of elements in one to one correspondence with ${\mathbf Z}$ (i.e.  the set of integers);
\item continuous groups: Lie groups for gauge theories, classification of particles, spin group, group of symmetry of space-time ( Poincar\' e group,..)
\end{itemize}
\end{itemize}

An example of Lie group:                                                                                       

 Group of rotations in the real plane around an origin O. It is then defined by $2 \times 2$ orthogonal matrices (with real entries):
$$ 
  R(\theta)= \left( \begin{array}{crcr}
\:\:\:  \cos \theta & -\sin \theta \:\:\: \\ \:\:\:  \sin \theta & \cos \theta  \:\:\:   \end{array}\right) 
 \;\;\;  \;\;\;  0 \leq \theta \leq 2 \pi
 $$
 such a matrix transforming the 2 dim. real vector  $\vec{X}= (x, y)$  into $R(\theta) \vec{X} =  \vec{X\prime}$ with components:
  \bea
  x\prime & =&  x \cos \theta- y \sin \theta \nonumber \\
   y\prime & = & x \sin \theta + y \cos \theta
   \eea
    such that :
    \be
\vec{X} \cdot  \vec{X\prime}= \vec{X\prime} \cdot \vec{X}  \;\;\;  \vec{X} \cdot  \vec{X \prime} = x x\prime + y y\prime   \;\;\;  \mbox{(norm conservation)}
    \ee
Let us recall that:
\be   
 \cos \theta =1- \frac{\theta^2}{2!} + \ldots;       \;\;\;  \;\;\; \sin \theta =  \theta -  \frac{\theta^3}{3!} + \ldots  
\ee
leading, after a simple computation, to : 
\be
       R(\theta)    = \mathbf{1} +\theta  \mathbf{M} + \ldots = \exp(\theta  \mathbf{M})  
\ee
where 
$$  \mathbf{M} = \left( \begin{array}{crcr}
\:\:\:  0 & - 1 \:\:\: \\ \:\:\: 1 &  0 \:\:\:   \end{array}\right) $$
\noindent $\mathbf{M}$ is called the infinitesimal generator of the rotation group in two dimension, itself usually denoted by $\mathcal{O}(2)$. $\theta$ is called the parameter of $\mathcal{O}(2)$. 
Actually, we have just considered one of the simplest Lie group: it has one and only one continuous parameter ($\theta$ determines completely the angle of rotation around the origin for the group element $R(\theta)$).

Let us make things more complicated, and consider now the group of two by two unitary matrices, that is the group of $2 \times 2$ complex matrices $\mathcal{U}$ satisfying
$$(\mathcal{U} \vec{X},\mathcal{U} \vec{X}) = (\vec{X}, \vec{X})$$ 
 but with the components of  $\vec{X}$  being complex numbers:                $ x= a + i b, y= c + i d$  and the scalar product $(\vec{X}, \vec{X})$ being defined by: 
$$(\vec{X}, \vec{X}) = (a- i b) \times  (a + i b) + (c- i d) \times (c + i d)$$
Then, the condition  $(\mathcal{U} \vec{X},\mathcal{U} \vec{X}) = (\vec{X}, \vec{X})$ will be satisfied, for any $\vec{X}$,  if and only if:    $\mathcal{U}^{\dag}  = \mathcal{U}^{-1}$  with  $\mathcal{U}^{\dag} $ obtained by replacing each entry of  $\mathcal{U}$  by its complex conjugate and transposing the matrix (with respect to the main diagonal) and $\mathcal{U}^{-1}$  denoting the inverse of $\mathcal{U}$ ($\mathcal{U}  \mathcal{U}^{-1} =  \mathbf{1} $, where   $\mathbf{1} $ is the  $2 \times 2$ identity matrix).
 
 \noindent {\bf Example}: 
 $$ 
\mathcal{U}= \left( \begin{array}{crcr}
\:\:\:  a+ib  \, & \, c+id  \:\:\: \\ \:\:\: e + if \,  & \,  g+ih  \:\:\:   \end{array}\right) 
 \;\;\;  \;\;\; \mathcal{U}^{\dag}= \left( \begin{array}{crcr}
\:\:\:  a- ib  \, &  \,  e-if  \:\:\: \\ \:\:\:  c-id \,  &  \,  g-ih  \:\:\:      \end{array}\right) 
 $$
 The usual notation for the group of unitary  $2 \times 2$ matrices is $\mathcal{U}(2)$, and is $\mathcal{SU}(2)$ for its subgroup with elements $\mathcal{U}$ of  $\mathcal{U}(2)$ satisfying det ($\mathcal{U}$)= 1.                                       As for the  $\mathcal{O}(2)$ case,  an exponential expression can be obtained for any $\mathcal{U}$ in $\mathcal{SU}(2)$:
 \be
\mathcal{U} = e^{i \left(a \sigma_1 + b \sigma_2 + c \sigma_3\right)}    \;\;\;    \;\;\;  \;\;\;  a, b, c \, \in \mathbf{R}
 \ee
 where the three $2 \times 2$ $\sigma$ matrices  are the so called Pauli matrices defined as follows:
  \[\sigma_{1} = \left( \begin{array}{crcr}
\:\:\:  0 & 1 \:\:\: \\ \:\:\: 1 & 0  \:\:\:   \end{array}\right) 
 \;\;\;\;\;\;\;  \sigma_{2} = \left( \begin{array} {crcr} 
\:\:\: 0 &  - i \:\:\: \\ \:\:\: i & 0   \:\:\:  \end{array}\right)
 \;\;\;\;\;\;\;  \sigma_{3} = \left( \begin{array} {crcr}
\:\:\: 1 & 0 \:\: \\ \:\:\:  0 & -1 \:\:\:  \end{array}\right)\]
                           and satisfying the commutation relations: 
\bea
&& [ \sigma_{1}, \, \sigma_{2}] =  \sigma_{1}  \sigma_{2} -  \sigma_{2} \sigma_{1} = 2 i \sigma_{3} \label{eq:com}\\
&&    [ \sigma_{2}, \, \sigma_{3}] = 2 i \sigma_{1}  \;\;\; [ \sigma_{3}, \, \sigma_{1}] = 2 i \sigma_{2} 
\eea

In $\mathcal{SU}(2)$   there are three real continuous parameters, namely a, b and c and then three infinitesimal generators:   $\sigma_1, \sigma_2, \sigma_3$.

In any Lie group, the infinitesimal generators form a basis of the Lie algebra of the corresponding Lie group.

Let us give the general definition of a Lie algebra \rm{A}:

\noindent {\bf Definition 5}: A  Lie algebra is an algebra, that is first a linear vector space on  $\bf{R}$ (or on $\bf{C} \dots$) with a second internal law @ satisfying (if we denote + the first law of the vector space):
\be
\forall \, x,y,z  \in \, {\rm A}\	, : \, (x+y)  @ z= x@z+y@z    \;\;\;    \;\;\;        x@(y+z)=x@z+y@z 
\ee
Moreover, this second law satisfies:   
\be
\forall \, x,y,z  \in \, {\rm A}  \, : \,x@y= -y@x    \;\;\;    \;\;\;     x@(y @z)+ y@(z @x) + z@(x @y) = 0 
\ee
One can easily note that, in our case, the @ internal law is just the commutator, denoted by [., .]:                                                  
  \be                                                 
 [x,\,y ]= x. y \, - \, y. x
 \ee

The property of a Lie group (or continuous group of transformations satisfying some more analytical properties that we will not consider here in order not to overload this first introductory lesson)element to be written as an exponential of an element of its Lie algebra is particularly useful. Indeed, it will now be possible to work most of the time (at least when topological questions are not on purpose) with the Lie algebra, that is replacing tedious and enormous computations on the group by computations involving mainly linear algebras! That will be particularly precious for constructing and studying representations of Lie groups.

Actually, the $\mathcal{SU}(2)$ group is also known as the  ``spin''  group in elementary particle physics.

 Now, let us consider  a little more to notion of representations, as mathematically defined by  Definition 3.
Indeed, the same group can act non trivially on spaces of different dimensions. The  $\mathcal{SU}(2)$ group is defined by the set of $2 \times 2$ unitary matrices: then its natural space of representation is the 2-dim. complex plane. We can say that the ``fundamental''  representation of $\mathcal{SU}(2)$ is given by the $2 \times 2$ unitary matrices of determinant = 1, acting on the 2 dimensional complex plane ${\bf C}_2$, which we call the ``representation''  space. 

We will now construct other $\mathcal{SU}(2)$ representations. But, before, let us remark that any element $\mathcal{U}= exp (a\sigma_1 + b\sigma_2 +c\sigma_3)$  in
 $\mathcal{SU}(2)$  can be transformed into another element of 
 $\mathcal{SU}(2) \; V = U \prime \; U \;  (U \prime)^{-1}$
 which is diagonal. Indeed,  the matrix $H=  a\sigma_1 + b\sigma_2 +c\sigma_3$ is Hermitian, that is  $H = H^{\dag}$ (see definition above) and so can be diagonalized by an unitary matrix $U \prime$ , whence V diagonal. Note that $U$ and $V$ are mathematically equivalent in the sense that there is a change of basis in ${\bf C}_2$ -  actually given by $U\prime$ - which will allow to see $U$ as a diagonal matrix.  And we will have:      
                               $$   V= exp ( d \sigma_3) \qquad \qquad       d \; real $$
At the Lie algebra level, the diagonal generator ($1/2 \sigma_3$) has two eigenvalues: +1/2 and -1/2 associated to the two eigenvectors (1, 0)  and (0,-1): for a physicist, these are the  two spin states $\ua$ and $\da$.  Moreover, one can see that the matrices:     $\sigma_{\pm} = \sigma_1 \pm i \sigma_2$
transforms the vector (0,1) into (1,0) and (1,0) into the null vector (0,0) (resp.  (1,0) into (0,1) and (0,1) into (0,0)) 
                          (they are called  ``raising and lowering operators'' ).   
We also note the commutation relations:
 \be
[\sigma_3, \, \sigma_{+}]= + 2 \sigma_{+}       \;\;\;    \;\;\;            [\sigma_{3}, \, \sigma_{-}]= + 2 \sigma_{-}     \;\;\;    \;\;\;        [\sigma_{+}, \, \sigma_{-}]= 4 \sigma_{3}
 \label{eq:crs}
 \ee
                                            But one knows that there are not only particles of spin 1/2, there are also particles of spin 0, 1, 3/2, 2,..They will lie in other representations of the group  $\mathcal{SU}(2)$. Let us see how to construct them.
For such a purpose we need to define the tensorial product of two vector spaces.

 \noindent {\bf Definition 6}: Let  {\bf V}  and 
  $\bf{V\prime}$ two vector spaces of respective dimensions $n$ and $n\prime$, and respective basis  ($\vec{e_1}, \ldots,  \vec{e_n}$) and  ($\vec{e_1}, \ldots,  \vec{e_n \prime}$). We define the tensor product  ${\bf V}  \otimes  {\bf V\prime}$ as the vector space of dimension $n  \times n\prime$ with elements: 
                                        $\sum_{(i,j)} \, \alpha(i,j)  [ \vec{e_i} \otimes \vec{e_j \prime}]$
with $ i= 1,\ldots,n$ and $j= 1,\ldots,n\prime$ and   $\alpha(i,j)  \,  \in {\bf R}$.
If the group  $\mathcal{G}$ acts on ${\bf V}$ via the representation $D$ and on  $\bf{V\prime}$  via the representation $D\prime$ such that  $\forall \, g \in \mathcal{G} \ra D(g)$ acts on  $\bf{V}$ and  $ \forall \, g \in \mathcal{G} \ra D\prime(g)$ acts on  $\bf{V\prime}$,                
   on  $\bf{V}  \otimes  \bf{V\prime}$ one can define the $\mathcal{ G}$-epresentation $D \otimes D\prime$ such that:  $\forall \, g \in \mathcal{G} \ra ( D \otimes D\prime)(g)=D(g) \otimes D\prime(g)$.

Now, let us take as {\bf V} the 2-dimensional complex vector space ${\bf C}_2$ and consider the action of $\mathcal{G} =\mathcal{SU}(2)$ on ${\bf V} \otimes {\bf V}$. Then any $U \, \in \mathcal{G}$ acts on any vector $ v \otimes v\prime$  of  ${\bf V} \otimes {\bf V}$ as: 
$$ \forall \, U \in \, \mathcal{G} \ra  U(v) \otimes U (v\prime)$$
But we know that we can rewrite U as: 
$$ U =   \exp({\bf M}) = \bf{1} + M + \ldots$$

   Then, infinitesimally, we will have:
   \be
 (\bf{1} + M + \ldots)(v) \otimes   (\bf{1} + M + \ldots)(v\prime) =  (\bf{1} \otimes \bf{1} + M \otimes \bf{1}  + \bf{1} \otimes M +  \ldots) (v \otimes v\prime)
   \ee
therefore, infinitesimally - or in other words: at the Lie algebra level, we will have:
   \be
                                       v \otimes v\prime \lra  M (v)\otimes v\prime + v \otimes M (v\prime)
   \ee
So, let us start with the vector:  $ \ua \otimes \ua$.
The $\sigma_{-}$-action  will give: 
   \be
                       ( \ua \otimes \ua)     \lra     (  \ua \otimes \da    +   \da \otimes \ua)  \lra     (\da \otimes \da ) \lra 0
   \ee
But we know that we have four vectors in the tensor space under consideration. After some simple computation, one can see that the vector:
                                             $(\ua \otimes \da  - \da \otimes \ua)$
 is such that the action of  $\sigma_{-}$  as well as the action of $\sigma_{+}$  on it gives 0.
Thus, the four dimensional space ${\bf C}_2  \otimes {\bf C}_2$ is indeed a good representation space for  $\mathcal{G} =\mathcal{SU}(2)$, but it splits into two subspaces , one of dimension 3 and one of dimension 1, each of them being a good representation of  $\mathcal{G}$. Actually, each of these two ${\bf C}_2  \otimes {\bf C}_2$ subspaces are  ``invariant subspaces under $\mathcal{G}$ '' , and we have obtained what are called ``irreducible representations of $\mathcal{G}$ ''. Let us define correctly these objects:

  \noindent {\bf Definition 7}:  Let D be a representation of the group  $\mathcal{G}$ in {\bf V}. the subspace        $\bf{E} \subset \bf{V}$ is an invariant subspace of {\bf V} under D if:
                                       $$ \forall \, g \in \mathcal{G} \,:     (D(g)) {\bf E} \subset {\bf E}$$

\noindent {\bf Definition 8}:  The representation D of $\mathcal{G}$ in V is irreducible if there is no invariant subspace, except the trivial one (i.e.: 0). If not D is said reducible.

In the just considered case, the representation 
  ${\bf C}_2 \otimes {\bf C}_2$  of $\mathcal{SU}(2)$, is reducible and decomposes in two separate representations, we have a ``partition''  of the space)) irreducible  $\mathcal{SU}(2)$ representations, of dimension 3 and 1 respectively. 
Remark: The basis $\sigma_{+}$, $\sigma_{-}$ and $\sigma_{3}$  is more likely associated to the Lie algebra of the group $\mathcal{S}l(2,R)$, often - and incorrectly -  written $\mathcal{S}l(2)$, and defined as the group of $2 \times 2$ real matrices with determinant = 1.  It is a ï¿½ï¿½non-compactï¿½ï¿½ form of $\mathcal{SU}(2)$; both $\mathcal{S}l(2)$, and $\mathcal{SU}(2)$ possess the same set of  irreducible finite dimensional representations.

{\bf  Conclusions from the above discussion}:
  \begin{itemize}
\item	from the $2$-dimensional representation of  $\mathcal{SU}(2)$ we have constructed the (irreducible) $3$-dimensional representation  and the 1-dimensional (or trivial)one. The first one corresponds to the  ``spin''  one representation, with the three states with eigenvalue 1, 0, and -1, while the second is the ``spin 0''  representation with only one state with 0 eigenvalue;
\item	we have also  ``appreciated''  the powerfulness of a Lie group to have a Lie algebra which allows easier computations (a Lie algebra satisfying the property of a linear algebra).
  \end{itemize}
{\bf More on (finite dimensional) representations of the  $\mathcal{SU}(2)$ group}:
 Any irreducible finite dimensional representation of  $\mathcal{SU}(2)$ is usually denoted $D(j)$ with $j$ being a positive (or null ) integer or half integer. The $D(j)$ representation contains $(2j+1)$ states, each state being an eigenstate of the generator  corresponding to  $\sigma_{3}$ with eigenvalue: $j, j-1, \ldots,-j+1,-j$ respectively. The ``spin 1''  representation above discussed is then D(1) with three states associated to the eigenvalue $ +1,0,-1$ respecively., while the ``spin 1/2''  representation $D(1/2)$ contains two states with  $\sigma_{3}$ eigenvalue +1/2 and ï¿½ 1/2.                                             
Let us add that the product $D(j) \otimes D(j\prime)$   of the two representations $D(j)$ and $D(j\prime)$ of  $\mathcal{SU}(2)$  decomposes as the sum of the irreducible representations:
  \be
                  D(j) \otimes D(j\prime) = D(j+j\prime) \oplus D(j+j\prime-1) \oplus  \ldots \oplus D(| j-j\prime|)
  \ee 

Finally, let us mention that the  main Lie groups, at least with the most  simple properties, are the following (they are called ``simple groups''    but we will not overload our text with the definition of a simple group)
  \begin{itemize}
\item $\mathcal{SO}(n)$: orthogonal groups in $n$-dimensional real space (i.e.  group of real  $n \times n$ orthogonal matrices);
  \item $\mathcal{SU}(n)$: unitary groups in $n$-dimensional complex space (i.e. group of
  $n \times n$ complex unitary matrices);
\item	$\mathcal{S}p(n)$: group of $2n \times 2n$ symplectic matrices. 
	\item there are also 5 ``exceptional''   groups: the word exceptional is used because they do not enter in infinite series as the above ones. 
\end{itemize}
More details and informations- and complete definitions -  on this section can be found in \cite{FSS}.
  
  \subsection{A II: Quantum Groups and Crystal basis }

It is of course not our purpose to develop in detail the theory of  quantum groups as it appeared in the works of V. Drinfeld on one hand and of M. Jimbo on the other hand in the middle of the eighties, but to provide the minimum of definitions and properties of a quantum group  $\mathcal{U}_q(g)$, g being  the Lie algebra of a Lie group  $\mathcal{G}$ and $q $ denoting the corresponding deformation parameter.
 We have noticed in the first part of the Appendix that a Lie algebra $G$  of a Lie group $\mathcal{G}$ has a structure of linear vector space. Let us consider now the universal enveloping algebra of this Lie algebra, i.e. the space of polynomials and formal power series in $g \in G$ on which we apply the commutation relations appropriate for that Lie algebra. Then the quantum group $\mathcal{U}_q(g)$ will be a deformation relative to the parameter $q$ of the universal enveloping algebra $g$. More explicitly, let us consider the example of the $\mathcal{U}_q((Sl(2))$ quantum group and let us denote  $J_{+},J_{-}$ and $J_3$ the generators corresponding   $\sigma_{+}$, $\sigma_{-}$ and $\sigma_{3}$ in the $2$-dimensional space representation, then we have:
\be
[ J_3, \,J_{\pm} ] = \pm  J_{\pm}   \;\;\;    \;\;\;             [J_{+},	\, J_{-}] =  \frac{q^{J_{3}} - q^{-J_{3}}}{q^{1/2} - q^{-1/2}}
\ee
We remark that when the parameter $q \to 1$, one recovers the  $\mathcal{S}l(2)$commutation relations\footnote{The commutation relations of eq.(\ref{eq:csl2}) follow from eq.(\ref{eq:crs}) defining $J_i = \sigma_i/2$, $i =1,2,3$.}
\be
[ J_3, \,J_{\pm} ] = \pm J_{\pm}   \;\;\;    \;\;\;             [J_{+},	\, J_{-}] =  2 J_{3}
\label{eq:csl2}
\ee
Another important limit is the one corresponding to $q \to 0$. A detailed study of this case has first been done by M. Kashiwara  \cite{K} who found  particularly well behaved base called`` Crystal Base'' .    Particularly interesting for our purpose is the rule providing the product of two irreducible representations of $\mathcal{U}_q(G)$ when $q \to 0$.
A remarkable property of such a rule is that the elements in the obtained representation spaces arising from the product of two representations are not linear combinations of states, as in the case of an usual group as  $\mathcal{S}l(2)$ (see the example considered in Appendix A I), but only made of a single product of the form  $u\prime \otimes v\prime$ with 
$u\prime \in \,  B_1$ and  $v\prime \in \, B_2$: see Theorem 2.1 in Subsection 2.1.

We can make more explicit the way of computing such states by considering the example of the product $D(3/2) \otimes D(1)$ which decomposes as  - see above in  Appendix AI:
                     $$D(3/2) \otimes  D(1) = D(5/2) \oplus D(3/2) \oplus D(1/2)  $$    
In Fig.1, we have representedby black points on an horizontal line the four $J_3$ eigenstates $3/2, 1/2, - 1/2, -3/2$ of $D(3/2$)   and on a vertical line the three $J_3$ eigenstates $1, 0,-1$ of $D(1)$. Using the above theorem, the six states in the obtained representation $D(5/2)$ appear as the black points in the upper elbow constituted by the two - one horizontal and one vertical -  segments, the same for $D(3/2)$ in the lower elbow, while  finally the two states of the $D(1/2$) representation states show up in the small horizontal segment below:

It is this property which is intensively used in our symmetry approach of the genetic code. In our model, the used quantum group is $\mathcal{U}_q (Sl(2)\oplus Sl(2))$, $q \to 0$.
Let us make more explicit the construction of dinuclotides from the product:
                    $$ (\frac{1}{2}, \frac{1}{2}) \otimes  (\frac{1}{2}, \frac{1}{2}) = (1,1) \oplus (0,1) \oplus (0,1) \oplus  (0,0)$$
and following the prescription given in the scheme(1)in Subsection 2.1. Starting from the state CC, the action of $J_{-}$ in $Sl(2)_H$  will provide UC and UU successively, while the action of $J_{-}$ in $Sl(2)_V$ gives GC and GG from CC,  AC and AG from UC, and finally AU and AA from UU. Using once more the diagrammatic rule above -  that is the Kashiwara theorem - one gets CU as a singlet of $Sl(2)_H$  but member of a triplet of $Sl(2)_V$ , in the same way CG as a singlet of $Sl(2)_V$ but in a triplet of $Sl(2)_H$  and finally CA as a singlet of both $Sl(2)$. \\

\section{Acknowledgments}

P.S. would like to express his gratitude to Professor R. MondainiÊ for his kind invitation to present our results at the BIOMAT 2016 International Symposium, Tianjin,China, and to encourage us to write a developed review on our model. 

\noindent He is also indebtedÊ to Professor B. Dragovich for his warmÊ invitation as a speaker toÊ the BelBI2016 International Symposium, Belgrade,Ê and also for Ê Ê Ê Ê Ê Ê Ê Ê Ê Ê Ê Ê Ê Ê Ê Ê Ê Ê Ê Ê Ê Ê Ê Ê Ê Ê Ê Ê Ê Ê Ê Ê Ê Ê Ê Ê Ê Ê Ê Ê Ê Ê Ê Ê Ê his constant and friendly support in the development of our model.

\begin{figure}[htbp]
\caption{Diagram of the tensor product of irreps. $D(3/2) \otimes D(1)$ in the Crystal Basis.}
\includegraphics[width=14cm]{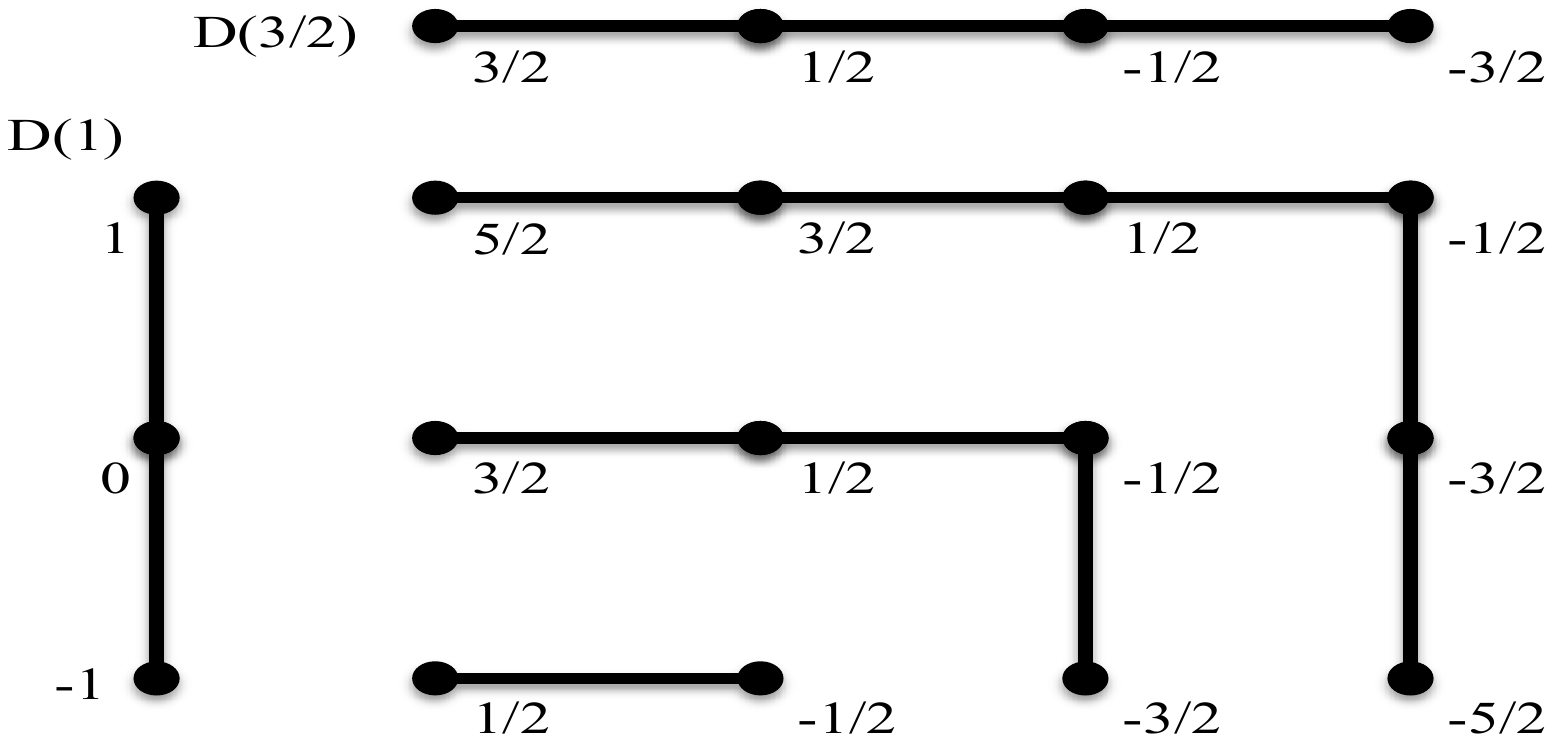}
 \label{TP}
 \end{figure}

\end{document}